\documentclass[11pt]{article}
\usepackage{amsmath}
\usepackage{xspace}
\usepackage{amssymb}
\usepackage{epsfig}

\newtheorem{Thm}{Theorem}
\newtheorem{Lem}[Thm]{Lemma}
\newtheorem{Cor}[Thm]{Corollary}
\newtheorem{Prop}[Thm]{Proposition}
\newtheorem{Claim}{Claim}
\newtheorem{Obs}{Observation}

\newtheorem{Def}{Definition}
\newenvironment{proof}{\noindent {\textbf{Proof }}}{$\Box$ \medskip}

\newcommand{\defeq}{\stackrel{\mathrm{def}}{=}}
\newcommand\B{\{0,1\}}

\newcommand\ket[1]{| #1 \rangle}

\newcommand\pr{\mbox{\bf Pr}}
\newcommand\av{\mbox{\bf{\bf E}}}

\newcommand {\ie} {\emph{i.e.}\xspace}
\newcommand {\st} {\emph{s.t.}\xspace}
\newcommand {\vs} {\emph{vs.}\xspace}

\begin{document}
\thispagestyle{empty}
\title{\textbf{New upper and lower bounds for
randomized and quantum Local Search}\thanks{This research was
supported in part by NSF grants CCR-0310466 and CCF-0426582.}}
\author{Shengyu Zhang \thanks{Computer Science Department, Princeton
University, NJ 08544, USA.  Email: szhang@cs.princeton.edu}}
\date{}
\maketitle

\abstract{Local Search problem, which finds a local minimum of a
black-box function on a given graph, is of both practical and
theoretical importance to combinatorial optimization, complexity
theory and many other areas in theoretical computer science. In this
paper, we study the problem in the randomized and quantum query
models and give new lower and upper bound techniques in both models.

The lower bound technique works for any graph that contains a
product graph as a subgraph. Applying it to the Boolean hypercube
$\B^n$ and the constant dimensional grids $[n]^d$, two particular
product graphs that recently drew much attention, we get the
following tight results:
\[
\begin{array}{ll}
\vspace{.5em} RLS(\B^n) = \Theta(2^{n/2}n^{1/2}), & QLS(\B^n) =
\Theta(2^{n/3}n^{1/6}), \\ RLS([n]^d) = \Theta(n^{d/2}) \text{ for }
d\geq 4, & QLS([n]^d) = \Theta(n^{d/3}) \text{ for } d\geq 6.
\end{array}
\]
Here $RLS(G)$ and $QLS(G)$ are the randomized and quantum query
complexities of Local Search on $G$, respectively. These improve the
previous results by Aaronson \cite{Aa04}, Ambainis (unpublished) and
Santha and Szegedy\cite{SS04a}.

Our new algorithms work well when the underlying graph expands
slowly. As an application to $[n]^2$, a new quantum algorithm using
$O(\sqrt{n}(\log\log n)^{1.5})$ queries is given. This improves the
previous best known upper bound of $O(n^{2/3})$ (Aaronson,
\cite{Aa04}), and implies that Local Search on grids exhibits
different properties in low dimensions.

\clearpage

%%%%%%%%%%%%%%%%%%%%%%%%%%%%%%%%%%%%%%%%%%%%%%%%%%%%%%%%%%%%%%%%%%%
%%%%%%%%%%%%%%%%%%%%%%%%%%%%%%%%%%%%%%%%%%%%%%%%%%%%%%%%%%%%%%%%%%%
\section{Introduction} \label{sec: introduction} Many important
combinatorial optimization problems arising in both theory and
practice are \textbf{NP}-hard, which forces people to resort to
heuristic searches in practice. One popular approach is Local
Search, by which one first defines a \emph{neighborhood structure},
then finds a solution that is locally optimal with respect to this
neighborhood structure. In the past two decades, Local Search
approach has been extensively developed and ``has reinforced its
position as a standard approach in combinatorial optimization" in
practice \cite{AHL+97}. Besides the practical applications, the
problem also has many connections to the complexity theory,
especially to the complexity classes \textbf{PLS}
\footnote{Polynomial Local Search, introduced by Johnson,
Papadimitriou, and Yannakakis \cite{JPY88}.} and \textbf{TFNP}
\footnote{The family of total function problems, introduced by
Megiddo and Papadimitriou \cite{MP91}.}. For example, the 2SAT-FLIP
problem is Local Search on the Boolean hypercube graph $\B^n$, with
the objective function being the sum of the weights of the clauses
that the truth assignment $x\in \B^n$ satisfies. This problem is
complete in \textbf{PLS}, implying that the Boolean hypercube $\B^n$
has a central position in the studies of Local Search. Local Search
is also related to physical systems including folding proteins and
to the quantum adiabatic algorithms \cite{Aa04}. We refer readers to
the papers \cite{Aa04,OPS04,SS04a} for more discussions and the book
\cite{AL97} for a comprehensive introduction.

Precisely, Local Search on an undirected graph $G = (V,E)$ is
defined as follows. Given a function $f:V\rightarrow \mathbb{N}$,
find a vertex $v\in V$ such that $f(v)\leq f(w)$ for all neighbors
$w$ of $v$. A class of \emph{generic algorithms} that has been
widely used in practice is as follows: we first set out with an
initial point $v\in V$, then repeatedly search a better/best
neighbor until it reaches a local minimum. Though empirically this
class of algorithms work very well in most applications, relatively
few theoretical results are known about how good the generic
algorithms are, especially for the randomized (and quantum)
algorithms.

Among models for the theoretical studies, the query model has drawn
much attention \cite{Aa04,Al83,AK93,LT93,LTT89,SS04a}. In this
model, $f$ is given by a black-box, \ie $f(v)$ can be accessed by
querying $v$. We only care about the number of queries made, and all
other computations are free. If we are allowed to toss coins to
decide the next query, then we have randomized query algorithm. If
we are allowed use quantum mechanics to query all the positions (and
get corresponding answers) in superposition, then we have quantum
query algorithms. The deterministic, randomized and quantum query
complexities, are the minimum number of queries needed to compute
the function by a deterministic, randomized and quantum query
algorithm, respectively. We use $RLS(G)$ and $QLS(G)$ to denote the
randomized and quantum query complexities of Local Search on graph
$G$, respectively. Previous upper bounds on a general $N$-vertex
graph $G$ are $RLS(G) = O(\sqrt{N\delta})$ by Aldous \cite{Al83} and
$QLS(G) = O(N^{1/3}\delta ^{1/6})$ by Aaronson \cite{Aa04}, where
$\delta$ is the maximum degree of $G$. Both algorithms actually fall
into the category of generic algorithms mentioned above, with the
initial point picked as a best one over a certain number of random
samples. Immediately, two questions can be asked:

\vspace{.5em} 1. On what graphs are these simple algorithms optimal?

\vspace{.5em} 2. For other graphs, what better algorithms can we
have?

\vspace{.5em}

\noindent Clearly the first one is a lower bound question and the
second one is an upper bound question.

Previously for lower bounds, Aaronson \cite{Aa04} showed the
following results on two special classes of graphs: the Boolean
hypercube $\B^n$ and the constant dimensional grid $[n]^d$:
\begin{align}
& RLS(\B^n) = \Omega(2^{n/2}/n^2), & & QLS(\B^n) =
\Omega(2^{n/4}/n); \\
& RLS([n]^d) = \Omega(n^{d/2-1}/\log n), & & QLS([n]^d) =
\Omega(n^{d/4-1/2}/\sqrt{\log n}).
\end{align}
It has also been shown that $QLS([n]^2) = \Omega(n^{1/4})$ by Santha
and Szegedy in \cite{SS04a}, besides their main result that the
deterministic and the quantum query complexities of Local Search on
any graph are polynomially related. However, the question

\vspace{.5em} 3. What are the final values of $QLS$ and $RLS$ on
$\B^n$ and $[n]^d$? \vspace{.5em}

\noindent remains an open problem, explicitly stated in an earlier
version of \cite{Aa04} and also (partially) in \cite{SS04a}.

In this paper, we answer questions 1 and 2 for large classes of
graphs by giving both new lower and upper bound techniques for
randomized and quantum query algorithms. As a consequence, we
completely solve the question 3, except for a few small $d$'s where
our new bounds also significantly improve the old ones.

Our lower bound technique works for any graph that contains a
product graph as a subgraph. For two graphs $G_1 = (V_1, E_1)$ and
$G_2 = (V_2, E_2)$, their product $G_1 \times G_2$ is the graph $G =
(V,E)$ where $V = V_1 \times V_2$ and
\begin{equation}
E = \{(v_1\otimes v_2,v_1'\otimes v_2): (v_1,v_1')\in E_1, v_2\in
V_2\} \cup \{(v_1\otimes v_2, v_1\otimes v_2'): (v_2,v_2')\in E_2,
v_1\in V_1\}
\end{equation}
We will also use the notion of random walk on graphs to state the
theorem. Given a graph $G = (V,E)$, a random walk is a mapping
$W:V\rightarrow 2^V$ where $W(u)\subseteq \{u\}\cup \{v: (u,v)\in
E\}$. Intuitively, at each step the random walk $W$ goes from the
current vertex $u$ to a uniformly random vertex in $W(u)$. The walk
$W$ is \emph{regular} if $|W(u)| = c$ for each $u\in V$. Denote by
$p(u,v,t)$ the probability that the random walk starting at $u$ is
at $v$ after exactly $t$ steps. Let $p_t = \max_{u,v} p(u,v,t)$. The
following theorem is a special case of the general one (Theorem
\ref{thm: general lb}) in Section \ref{sec: general lb}.

\begin{Thm}\label{thm: special walk lb}
Suppose $G$ contains the product graph $G_1 \times G_2$ as a
subgraph, and $L$ is the length of the longest self-avoiding path in
$G_2$. Let $T = \lfloor L/2 \rfloor$, then for any regular random
walk $W$ on $G_1$, we have
\[
RLS(G) = \Omega\left(\frac{T}{\sum_{t=1}^{T} p_t}\right), \quad
QLS(G) = \Omega\left(\frac{T}{\sum_{t=1}^{T} \sqrt{p_t}}\right).
\]
\end{Thm}

The proof uses the quantum adversary method, which was originally
proposed by Ambainis \cite{Am00} and later generalized in different
ways \cite{Am03,BSS03,LM04,Zh04}. Recently Spalek and Szegedy made
the picture clear by showing that all these generalizations are
equivalent in power \cite{SS04b}. On the other hand, in proving a
particular problem, some of the methods might be easier to apply
than the others. In our case, the technique in \cite{Zh04}, which
generalizes the one in \cite{Am03} slightly in the form though,
turns out to work very well. Our proofs for the randomized lower
bounds will use the relational adversary method, which was proposed
by Aaronson \cite{Aa04} inspired by the quantum adversary method.

Both the quantum adversary method and the relational adversary
method are parameterized by input sets and weight functions of input
pairs. While previous works \cite{Aa04, SS04a} also use random walks
on graphs, a key innovation that distinguishes our work from the
previous ones and yields better lower bounds is that we decompose
the graph into two parts, the tensor product of which is the
original graph. We perform the random walk only in one part, and
perform a simple one-way walk in a self-avoiding path in the other
part, which serves as a ``clock" to record the number of steps taken
by the random walk in the first part. The tensor product of these
two walks is a random path in the original graph. A big advantage of
adding a clock is that the ``passing probability", the probability
that the random path \emph{passes} a vertex $v$ \emph{within} $T$
steps, is now the ``hitting probability", the probability that the
random walk in the first graph \emph{hits} $v$ \emph{after} exactly
$t$ steps, because the time elapses one-way and never comes back.
The fact that the hitting probability is much smaller than the
passing probability enables us to achieve the better lower bounds.
Another advantage of the clock is that since the walk in the second
part is self-avoiding, the resulting random path in the original
graph is self-avoiding too, which makes the analysis much easier.

Applying it to the two graphs $\B^n$ and $[n]^d$, we improve
previous results and show tight bounds on both $RLS$ and $QLS$
(except for a few cases in the low dimensional grids).
\begin{Thm} \label{thm: hypercube}
\begin{equation}
RLS(\B^n) = \Theta(2^{n/2}n^{1/2}), \qquad QLS(\B^n) =
\Theta(2^{n/3}n^{1/6}).
\end{equation}
\end{Thm}

\begin{Thm} \label{thm: grid lb}
\begin{equation}
RLS([n]^d) =
\begin{cases}
\Theta(n^{d/2}) & \text{if } d \geq 4, \\
\Omega((n^3/\log n)^{1/2}) & \text{if } d = 3, \\
\Omega(n^{2/3}) & \text{if } d = 2.
\end{cases} \quad
QLS([n]^d) =
\begin{cases}
\Theta(n^{d/3}) & \text{if } d \geq 6, \\
\Omega((n^5/\log n)^{1/3}) & \text{if } d = 5, \\
\Omega(n^{6/5}) & \text{if } d = 4, \\
\Omega(n^{3/4}) & \text{if } d  = 3, \\
\Omega(n^{2/5}) & \text{if } d  = 2.
\end{cases}
\end{equation}
\end{Thm}
It is worth to note that to apply Theorem \ref{thm: special walk
lb}, we need not only know the mixing time of the random walk in
$G_1$, but also know \emph{its behavior before mixing}. So the
applications are not simply using standard upper bounds on the
mixing times, but involving heavy analysis on the whole mixing
processes.

When proving Theorem \ref{thm: grid lb} by Theorem \ref{thm: special
walk lb}, one difficulty arises: to decompose the grid $[n]^d$ into
two parts $[n]^m$ and $[n]^{d-m}$, we implicitly require that $m$ is
an integer. This lets us get lower bounds weaker than Theorem
\ref{thm: grid lb}, especially for low dimension cases. We get
around this problem by cutting one of the $m$ dimensions into many
blocks, and use different block to distinguish different time
windows. Between adjacent blocks are pairwise disjoint path
segments, which thus thread all the blocks into a very long one.
Using this technique, we can apply Theorem \ref{thm: special walk
lb} for any read-number dimension $m \leq d-1$.

\vspace{.5em} In the second part of the paper, we consider upper
bounds for Local Search. While the generic algorithms \cite{Aa04,
Al83} are simple and proven to be optimal for many graphs such as
the ones mentioned above, they are far from optimal for some other
graphs. For example, it is not hard to see an $O(\log N)$
\emph{deterministic} algorithm for the line graph $G$. Therefore, a
natural question is to characterize those graphs on which Local
Search is easy. It turns out that the expansion speed plays a key
role. For a graph $G = (V,E)$, the distance $|u-v|$ between two
vertices $u$ and $v$ is the length of the shortest path connecting
them. (Here the length of a path is the number of edges on the
path.) Let $c(k) = \max_{v\in V}|\{u: |u-v|\leq k\}|$. Apparently,
the smaller $c(k)$ is, the more slowly the graph expands. (Actually
$c(k)$ is an upper bound of the standard definition of the expanding
speed.) As a special case of Theorem \ref{thm: upper bound} in
Section \ref{sec: upper bound}, the following upper bounds in terms
of $c(k)$ hold.

\begin{Thm}
If \ $c(k) = O(k^\alpha)$ for some constant $\alpha \geq 1$, then
\begin{eqnarray}
RLS(G) =
\begin{cases} O\left( d^{\alpha-1} \log \log d\right)
& \text{ if } \alpha > 1, \\ O(\log d \log \log d) & \text{ if }
\alpha = 1.
\end{cases}
\qquad QLS(G) =
\begin{cases} O\left(
d^{\frac{\alpha-1}{2}} (\log \log d)^{1.5} \right) & \text{ if }
\alpha > 1, \\ O(\log d \log \log d) & \text{ if } \alpha = 1.
\end{cases}
\end{eqnarray}
where $d$ is the diameter of the graph $G$.
\end{Thm}
As a special case, on the line graph we get $\alpha = 1$ and hence
$RLS = O(\log n\log \log n)$, which helps to explain why Local
Search on the line graph is easy. Also, it immediately gives a new
upper bound for $QLS([n]^2)$ as follows. Together with Theorem
\ref{thm: grid lb}, this implies that Local Search on grids exhibits
different properties in low dimensions.

\begin{Thm}\label{thm: grid ub}
$QLS([n]^2) = O(\sqrt{n}(\log \log n)^{1.5})$
\end{Thm}

\noindent\emph{Other related results}. After the preliminary version
of this paper appeared, Verhoeven independently showed an upper
bound in terms of the genus of the graph \cite{Ve06}, giving an
$O(\sqrt{n}\log\log n)$ quantum algorithm for $[n]^2$. There is also
an unpublished result on $QLS(\B^n)$: it is mentioned in \cite{Aa04}
that Ambainis showed $QLS(\B^n) = \Omega(2^{n/3}/n^{O(1)})$.
\footnote{Another unpublished result was mentioned in \cite{SS04a}
that Verhoeven showed $RLS([n]^2) = \Omega(n^{1-\delta})$ for any
constant $\delta>0$. But according to an author of \cite{SS04a}, the
proof was never written up and this question should be considered
now to be still open.}

%%%%%%%%%%%%%%%%%%%%%%%%%%%%%%%%%%%%%%%%%%%%%%%%%%%%%%%%%%%%%%%%%%%
%%%%%%%%%%%%%%%%%%%%%%%%%%%%%%%%%%%%%%%%%%%%%%%%%%%%%%%%%%%%%%%%%%%
\section{Preliminaries and notations}
We use $[M]$ to denote the set $\{1, 2, ..., M\}$. %Define the sign
%function to be $sign(z) = 1$ if $z> 0$, $-1$ if $z< 0$ and 0 if
%$z=0$.
For an $n$-bit binary string $x = x_0...x_{n-1}\in \B^n$, let
$x^{(i)} = x_0...x_{i-1}(1-x_i)x_{i+1}...x_{n-1}$ be the string
obtained by flipping the coordinate $i$.

For graphs $G_1 = (V_1,E_1)$ and $G_2= (V_2,E_2)$, we say that $G_1$
is a subgraph of $G_2$ if $V_1\subseteq V_2$ and $E_1\subseteq E_2$.
Apparently, any local optimum in $G_2$ is also a local optimum in
$G_1$ (but not the other way around in general), therefore any lower
bound for $G_1$ is also a lower bound for $G_2$.

We will use $v_1 \otimes v_2$ to range over the set $V_1 \times
V_2$. There are various ways to define a product graph $G_1 \times
G_2 = (V_1 \times V_2, E)$ by different choices of $E$. Three
possibilities are
\begin{enumerate}
\item $E = \{(v_1\otimes v_2,v_1'\otimes v_2): (v_1,v_1')\in E_1, v_2\in V_2\}
\cup \{(v_1\otimes v_2, v_1\otimes v_2'): (v_2,v_2')\in E_2, v_1\in
V_1\}$;

\item $E' = \{(v_1\otimes  v_2, v_1'\otimes v_2'): (v_1, v_1')\in E_1\cup
I_{V_1} \text{ and } (v_2, v_2') \in E_2\cup I_{V_2}\} -
I_{V_1\times V_2}$, where $I_{V} = \{(v,v): v\in V\}$;

\item $E'' = \{(v_1\otimes  v_2, v_1'\otimes  v_2'): (v_1, v_1')\in E_1\cup
I_{V_1} \text{ or } (v_2, v_2')\in E_2\cup I_{V_2}\} - I_{V_1\times
V_2}$.
\end{enumerate}
It is clear that $E\subseteq E' \subseteq E''$, and our lower bound
theorem will use the first definition $E$, making the theorem as
general as possible.

A path $X$ in a graph $G = (V,E)$ is a sequence $(v_1, ..., v_l)$ of
vertices such that for any pair $(v_i,v_{i+1})$ of vertices, either
$v_i = v_{i+1}$ or $(v_i,v_{i+1})\in E$. We use $set(X)$ to denote
the set of distinct vertices on path $X$. A path is self-avoiding if
$v_1$, ..., $v_l$ are all distinct. The length of a path $(v_1, ...,
v_l)$ is $l-1$. For two vertices $u,v\in V$, the distance $|u- v|_G$
is the length of a shortest path from $u$ to $v$. The subscript $G$
may be omitted if no confusion is caused.

The $(k,l)$-hypercube $G_{k,l} = (V,E)$ where $V=[k]^l$ and whose
edge set is $E = \{(u,v): \exists i\in \{0,...,l-1\}, \ \st\ |u_i -
v_i| = 1, \text{ and }u_j = v_j, \ \forall j\neq i\}$. Sometimes we
abuse the notation by using $[k]^l$ to denote $G_{k,l}$. Note that
both the Boolean hypercube and the constant dimension grid are
special hypercubes.\footnote{Here we identify the Boolean hypercube
$\B^n$ and $G_{2,n}$ since they are isomorphic.}

In an $N$-vertex graph $G = (V,E)$, a Hamilton path is a path $X =
(v_1, ..., v_{N})$ such that $(v_i, v_{i+1})\in E$ for any $i\in
[N-1]$ and $set(X) = V$. It is easy to check by induction that every
hypercube $[k]^l$ has a Hamilton path. Actually, for $l=1$, $[k]$
has a Hamilton path $(1,...,k)$. Now suppose $[k]^l$ has a Hamilton
path $P$, then a Hamilton path for $[k]^{l+1}$ can be constructed as
follows. First fix the last coordinate to be 1 and go through $P$,
then change the last coordinate to be 2 and go through $P$ in the
reverse order, and then change the last coordinate to be 3 and go
through $P$, and so on. For each $(k,l)$, let $HamPath_{k,l}=(v_1,
..., v_{N})$ be the Hamilton path constructed as above (where $N =
k^l$), and we define the successor function $H_{k,l}(v_i) = v_{i+1}$
for $i\in [N-1]$.

As mentioned in Section \ref{sec: introduction}, a deterministic
query algorithm for a function $f:I^n\rightarrow [M]$ accesses the
input $x\in I^n$ only by making queries in the form of ``$x_i = ?$".
Each query has cost 1, and all the other computation between queries
are free. A randomized query algorithm is the same except that the
algorithm can toss coins to decide which variable $x_i$ to ask next.
The quantum query model, formally introduced in \cite{BBC+01}, has a
working state in the form of
$\sum_{i,a,z}\alpha_{i,a,z}\ket{i,a,z}$. A quantum query on the
input $x$ proceeds as follows.
\begin{equation}
\sum_{i,a,z}\alpha_{i,a,z}\ket{i,a,z} \rightarrow
\sum_{i,a,z}\alpha_{i,a,z}\ket{i,a\oplus x_i,z}
\end{equation}
A $T$-query quantum query algorithm works as a sequence of
operations
\begin{equation}
U_0 \rightarrow O_x \rightarrow U_1 \rightarrow O_x \rightarrow ...
\rightarrow U_{T-1} \rightarrow O_x \rightarrow U_T
\end{equation}
where $O_x$ is as defined above, and each $U_t$ does not depend on
the input $x$. In both randomized and quantum query models, we can
allow a double-sided small constant error probability. The
deterministic, randomized and quantum query complexities, denoted by
$D(f)$, $R_2(f)$ and $Q_2(f)$, are the minimum numbers of queries we
need to make in order to compute the function by a deterministic,
randomized and quantum query algorithm, respectively. For more
details on the query models and the corresponding query
complexities, we refer to \cite{BW02} as an excellent survey.

%%%%%%%%%%%%%%%%%%%%%%%%%%%%%%%%%%%%%%%%%%%%%%%%%%%%%%%%%%%%%%%%%%%
\subsection{One quantum adversary method and the relational adversary method}
The quantum adversary method is one of the two powerful tools to
prove lower bounds on quantum query complexity; see \cite{HS05} for
an comprehensive survey of this research area. In this paper, we
will use the quantum adversary method proposed in \cite{Zh04}. The
definition and theorem given here are a little more general than the
original ones, but the proof remains unchanged.

\begin{Def}\label{weight scheme} Let $F: I^N\rightarrow [M]$ be an $N$-variate
function. Let $R\subseteq I^N\times I^N$ be a relation such that
$F(x) \neq F(y)$ for any $(x,y)\in R$. A weight scheme consists of
three weight functions $w(x,y)>0$, $u(x,y,i)>0$ and $v(x,y,i)>0$
satisfying $u(x,y,i)v(x,y,i) \geq w^2(x,y)$ for all $(x,y)\in R$ and
$i\in [N]$ with $x_i\neq y_i$. We further put
\begin{alignat}{2}
w_x & = \sum_{y': (x,y')\in R} w(x,y'), & \qquad w_y & = \sum_{x':
(x',y)\in R} w(x',y) \\
u_{x,i} & =  \sum_{y':(x,y')\in R, x_i \neq y'_i} u(x,y',i), &
\qquad v_{y,i} &= \sum_{x':(x',y)\in R, x'_i \neq y_i} v(x',y,i).
\end{alignat}
\end{Def}

\begin{Thm} \label{thm: Alb4} \emph{[Zhang, \cite{Zh04}]} For any
$F, R$ and any weight scheme $w,u,v$ as in Definition
\ref{weight scheme}, we have
\begin{equation}\label{eq: Alb4}
Q_2(F) =\Omega\left(\min_{(x,y)\in R,i\in [N]:\ x_i\neq y_i}
\sqrt{\frac{w_xw_y}{u_{x,i}v_{y,i}}}\right)
\end{equation}
\end{Thm}

In \cite{Aa04}, Aaronson gives a nice technique to get a lower bound
for randomized query complexity. We restate it using a similar
language of Theorem \ref{thm: Alb4}.

\begin{Thm} \label{thm: randomized lb} \emph{[Aaronson, \cite{Aa04}]}
Let $F: I^N\rightarrow [M]$ be an $N$-variate function. Let
$R\subseteq I^N\times I^N$ be a relation such that $F(x) \neq F(y)$
for any $(x,y)\in R$. For any weight function $w:R\rightarrow
\mathbb{R}^+$, we have
\begin{equation}\label{eq: relational adversary method}
R_2(F) =\Omega\left(\min_{(x,y)\in R,i\in [N], x_i\neq y_i}
\max\left\{\frac{w_x}{w_{x,i}},\frac{w_y}{w_{y,i}}\right\}\right)
\end{equation}
where
\begin{equation}
w_{x,i} =  \sum_{y':(x,y')\in R, x_i \neq y'_i} w(x,y'), \qquad
w_{y,i} = \sum_{x':(x',y)\in R, x'_i \neq y_i} w(x',y).
\end{equation}
\end{Thm}

Note that we can think of Theorem \ref{thm: randomized lb} as having
a weight scheme too, but requiring that $u(x,y,i) = v(x,y,i) =
w(x,y)$. This simple observation is used in the proof of Theorem
\ref{thm: hypercube} and \ref{thm: grid lb}.

%%%%%%%%%%%%%%%%%%%%%%%%%%%%%%%%%%%%%%%%%%%%%%%%%%%%%%%%%%%%%%%%%%%
%%%%%%%%%%%%%%%%%%%%%%%%%%%%%%%%%%%%%%%%%%%%%%%%%%%%%%%%%%%%%%%%%%%
\section{Lower bounds for Local Search on product graphs}\label{sec:
general lb}

In this section we prove a theorem which is stronger than Theorem
\ref{thm: special walk lb} due to a relaxation on the conditions of
the random walk. Suppose we are given a graph $G = (V,E)$, a
starting vertex $v_0$ and an assignment $W:V\times \mathbb{N}
\rightarrow 2^V$ \st for each $u\in V$ and $t\in \mathbb{N}$, it
holds that $W(u,t)\subseteq \{u\}\cup \{v: (u,v)\in E\}$ and that
$|W(u,t)| = c_t$ for some function $c$ of $t$. Intuitively, $W$
gives the candidates that the walk goes to for the next step, and
the random walk $(G, v_0, W)$ on graph $G$ proceeds as follows. It
starts at $v_0$, and at step $t\in \mathbb{N}$, it goes from the
current vertex $v_{t-1}$ to a uniformly random vertex in
$W(v_{t-1},t)$. We say \emph{a path $(v_0, v_1, ..., v_T)$ is
generated by the random walk} if $v_t\in W(v_{t-1}, t)$ for all $t
\in [T]$. Denote by $p(u,t_1, v,t_2)$ the probability that the
random walk is at $v$ after step $t_2$ under the condition that the
walk is at $u$ after step $t_1$. Let $p_t = \max_{u,v,t_1,t_2: \
t_2-t_1 = t} p(u,t_1,v,t_2)$. For $(u,u')\in E$, let
$q(u,u',t_1,v,t_2)$ be the probability that the walk is at $v$ after
step $t_2$, under the conditions that 1) the walk is at $u$ after
step $t_1$, and 2) the walk does not go to $u'$ at step $t_1+1$. The
following lemma on the relation of the two probabilities is obvious.

\begin{Lem}\label{lem: two prob}
If $|W(u,t_1+1)|>1$, then $q(u,u',t_1,v,t_2) \leq 2p(u,t_1,v,t_2)$.
\end{Lem}
\begin{proof} By considering the two cases of the step
$t_1+1$ (going to $u'$ or not), we have
\begin{equation}
p(u,t_1,v,t_2) = \frac{1}{|W(u,t_1+1)|}p(u',t_1+1,v,t_2) +
\left(1-\frac{1}{|W(u,t_1+1)|}\right)q(u,u',t_1,v,t_2).
\end{equation}
Thus
\begin{equation}
q(u,u',t_1,v,t_2) \leq
p(u,t_1,v,t_2)/\left(1-\frac{1}{|W(u,t_1+1)|}\right)\leq
2p(u,t_1,v,t_2).
\end{equation}
\end{proof}
\begin{Thm}\label{thm: general lb}
Suppose $G$ contains $G^w \times G^c$ as a subgraph, and $L$ is the
length of the longest self-avoiding path in $G^c$. Let $T = \lfloor
L/2 \rfloor$, then for the random walk $(G^w, v_0^w, W)$ on $G^w$,
we have
\begin{equation}
RLS(G) = \Omega\left(\frac{T}{\sum_{t=1}^{T} p_t}\right), \qquad
QLS(G) = \Omega\left(\frac{T}{\sum_{t=1}^{T} \sqrt{p_t}}\right).
\end{equation}
\end{Thm}

\begin{proof} Without loss of generality, we assume $G = G^w \times G^c$,
as Local Search on a subgraph is no harder than Local Search on the
original graph. We shall construct a random walk on $G$ by the
random walk $(G^w, v_0^w, W)$ on $G^w$ and a simple one-way walk on
$G^c$. Starting from some fixed vertex in $G$, the walk is proceeded
by one step of walk in $G^w$ followed by two steps of walk in $G^c$.
(We perform \emph{two} steps of walk in $G^c$ mainly for some
technical reasons, and this is where the factor of 2 in definition
$T = \lfloor L/2 \rfloor$ comes from.) Precisely, fix a
self-avoiding path $(z_{0,0}^c, z_{1,0}^c, z_{1,1}^c, z_{2,1}^c,
z_{2,2}^c, ..., z_{T,T-1}^c, z_{T,T}^c)$ of length $2T$ in $G^c$.
Let the set $P$ contain all the paths $X = (x_0^w\otimes z_{0,0}^c,
x_1^w\otimes z_{0,0}^c, x_1^w\otimes z_{1,0}^c, x_1^w\otimes
z_{1,1}^c, ..., x_{T}^w\otimes z_{T-1,T-1}^c, x_{T}^w\otimes
z_{T,T-1}^c, x_{T}^w\otimes z_{T,T}^c)$ in $G$ such that $x_0^w =
v_0^w$ and $(x_0^w, x_1^w, ..., x_T^w)$ is a path generated by the
random walk $(G^w, v_0^w, W)$. Define a problem \textsc{Path$_P$}:
given a path $X \in P$, find the end point $x_T^w \otimes
z_{T,T}^c$. To access $X$, we can ask whether $v\in set(X)$ for any
vertex $v \in V$, and an oracle $O$ will give us the Yes/No
answer.\footnote{Note that it is actually an oracle for the
following function $g: \B^n\rightarrow \B$, with $g(x) = 1$ if and
only if $x\in set(X)$. So strictly speaking, an input of
\textsc{Path$_P$} should be specified as $set(X)$ rather than $X$,
because in general, it is possible that $X\neq Y$ but $set(X) =
set(Y)$. For our problem, however, it is easy to check that for any
$X,Y\in P$, it holds that $X = Y \Leftrightarrow set(X) = set(Y)$.
Actually, if $X\neq Y$, suppose the first diverging place is $k$,
\ie $x_{k-1}^w = y_{k-1}^w$, but $x_k^w \neq y_k^w$. Then $Y$ will
never pass $x_{k}^w\otimes z_{k,k-1}^c$ because the clock
immediately ticks and the time always advances forward. (Or more
rigorously, the only point that $Y$ passes through $z_{k,k-1}^c$ is
$y_{k}^w \otimes z_{k,k-1}^c$. Since $y_{k}^w \neq x_k^w$,
$x_{k}^w\otimes z_{k,k-1}^c\notin set(Y)$.)} The following claim
says that the \textsc{Path$_P$} problem is not much harder than
Local Search problem.
\begin{Claim}\label{clm: reduction}
$R_2(\textsc{Path$_P$})\leq 2 RLS(G)$,\quad
$Q_2(\textsc{Path$_P$})\leq 2 QLS(G)$.
\end{Claim}
\begin{proof}
Suppose we have an $Q$-query randomized or quantum algorithm
$\mathcal{A}$ for Local Search, we shall give a $2Q$ corresponding
algorithm $\mathcal{B}$ for \textsc{Path$_P$}. For any path $X\in
P$, we define a function $f_{X}$ essentially in the same way as
Aaronson did in \cite{Aa04}: for each vertex $v\in G$, let
\begin{equation}
f_X(v) =
\begin{cases}
|v - x_0^w\otimes z_{0,0}^c|_G + 3T & \text{ if } v \notin set(X) \\
3(T-k) & \text{ if } v = x_{k}^w \otimes z_{k,k}^c \\
3(T-k)-1 & \text{ if } v = x_{k+1}^w \otimes z_{k,k}^c \neq x_{k}^w \otimes z_{k,k}^c\\
3(T-k)-2 & \text{ if } v = x_{k+1}^w \otimes z_{k+1,k}^c
\end{cases}
\end{equation}
It is easy to verify that the only local minimum is $x_T^w \otimes
z_{T,T}^c$.

Given an oracle $O$ and an input $X$ of the \textsc{Path} problem,
$\mathcal{B}$ simulates $\mathcal{A}$ to find the local minimum of
$f_X$, which is also the end point of $X$. Whenever $\mathcal{A}$
needs to make a query on $v$ to get $f_X(v)$, $\mathcal{B}$ asks $O$
whether $v\in set(X)$. If $v\notin set(X)$, then $f_X(v) = |v -
x_0^w\otimes z_{0,0}^c|_G + 3T$; otherwise, $v = x^w\otimes
z_{k+1,k}^c$ or $v = x^w\otimes z_{k,k}^c$ for some $x^w\in V^w$ and
$k$. Note that $k$ is known for any given vertex $v$. So if $v =
x^w\otimes z_{k+1,k}^c$, then $x^w = x_{k+1}^w$ and thus $f_X(v) =
3(T-k)-2$. Now consider the case that $v = x^w\otimes z_{k,k}^c$. If
$k=0$, then let $f_X(v) = 3T$ if $v = x_{0}^w \otimes z_{0,0}^c$ and
$f_X(v) = 3T-1$ otherwise. If $k \geq 1$, then $\mathcal{B}$ asks
$O$ whether $x^w\otimes z_{k,k-1}^c\in set(X)$. If yes, then $v =
x_{k}^w\otimes z_{k,k}^c$ and thus $f_X(v) = 3(T-k)$; if no, then $v
= x_{k+1}^w\otimes z_{k,k}^c\neq x_{k}^w\otimes z_{k,k}^c$ and thus
$f_X(v) = 3(T-k)-1$. Therefore, at most 2 queries on $O$ can
simulate one query on $f_X$, so we have a $2Q$ algorithm for
\textsc{Path$_P$} in both randomized and quantum cases.
\end{proof}

\noindent (Continue the proof of Theorem \ref{thm: general lb}) By
the claim, it is sufficient to prove lower bounds for
\textsc{Path$_P$}. We define a relation $R_P$ as follows.
\begin{equation}\label{eq: relation}
R_P = \{(X,Y): X\in P,\ Y\in P,\ X \text{ and } Y \text{ has
different end points}\}.
\end{equation}
For any pair $(X,Y)\in R_P$, where $X = (x_0^w\otimes z_{0,0}^c,
x_1^w\otimes z_{0,0}^c, x_1^w\otimes z_{1,0}^c, x_1^w\otimes
z_{1,1}^c, ..., x_{T}^w\otimes z_{T-1,T-1}^c, x_{T}^w\otimes
z_{T,T-1}^c, x_{T}^w\otimes z_{T,T}^c)$ and $Y = (y_0^w\otimes
z_{0,0}^c, y_1^w\otimes z_{0,0}^c, y_1^w\otimes z_{1,0}^c,
y_1^w\otimes z_{1,1}^c, ..., y_{T}^w\otimes z_{T-1,T-1}^c,
y_{T}^w\otimes z_{T,T-1}^c, y_{T}^w\otimes z_{T,T}^c)$, we write
$X\wedge Y = k$ if $x_0^w = y_0^w$, ..., $x_{k-1}^w = y_{k-1}^w$ but
$x_k^w \neq y_k^w$. Intuitively, $X\wedge Y = k$ if $k$ is the place
that the paths $X$ and $Y$ diverge for the first time. Note that if
$X\wedge Y = k$, then $x_k^w, y_k^w\in W(x_{k-1}^w, k)$ and thus
$|W(x_{k-1}^w, k)| \geq 2$. By Lemma \ref{lem: two prob}, this
implies that $q(x_{k-1}^w, x_k^w, k-1, v^w, j)\leq 2p_{j-k+1}$.

We choose the weight functions in Theorem \ref{thm: Alb4} by letting
\begin{eqnarray}\label{eq: def w}
w(X,Y) & = & 1/|\{Y'\in P: Y'\wedge X = k\}| \\
& = & 1/|\{X'\in P: X'\wedge Y = k\}| \\
& = & 1/[(c_k-1)c_{k+1}...c_T].
\end{eqnarray}
To calculate $w_{X} = \sum_{Y':(X,Y')\in R_P}w(X,Y')$, we group
those $Y'$ that diverge from $X$ at the same place $k'$:
\begin{eqnarray}
w_{X} & = & \sum_{k' = 1}^{T}\sum_{\scriptstyle \ Y':(X,Y')\in R_P
\atop \scriptstyle X\wedge Y' = k'}w(X,Y') \\
& = & \sum_{k' = 1}^{T}\sum_{\scriptstyle \ Y':(X,Y')\in R_P\atop
\scriptstyle X\wedge Y' = k'}\frac{1}{|\{Y'\in P: Y'\wedge X
= k'\}|} \\
\label{eq: prob} & = & \sum_{k' = 1}^{T} \pr_{Y'}[(X,Y')\in R_P|Y'
\wedge X =
k'] \\
& = & \sum_{k' = 1}^{T} \pr_{Y'}[(y')_{T}^w \neq x_{T}^w|Y'\wedge X
= k']
\end{eqnarray}
Here Equality \eqref{eq: prob} holds because all paths diverging
from $X$ firstly at $k'$ have the same probability
$1/[(c_{k'}-1)c_{k'}...c_T]$. Also note that the probability in the
last equality is nothing but $1-q(x_{k'-1}^w, x_{k'}^w, k'-1, x_T^w,
T)$, which is at least $1-2p_{T-k'+1}$. So we have
\begin{equation}
w_{X} \geq T-2\sum_{k'=1}^T p_{T-k'+1} = T-2\sum_{t=1}^{T} p_{t}.
\end{equation}
And similarly, we have $w_Y \geq T - 2\sum_{t=1}^{T} p_{t}$ too.

Now we define $u(X,Y,i)$ and $v(X,Y,i)$, where $i$ is a point
$x_{j+r}^w\otimes z_{j+s,j}^c\in set(X)-set(Y)$ or $y_{j+r}^w\otimes
z_{j+s,j}^c \in set(Y)-set(X)$. Here $(r,s)\in \{(0,0), (1,0),
(1,1)\}$, and $0\leq j\leq j+r \leq T$. Let
\begin{equation}\label{eq: def u}
u(X,Y,x_{j+r}^w\otimes z_{j+s,j}^c) = a_{k,j,r,s}w(X,Y), \quad
u(X,Y,y_{j+r}^w\otimes z_{j+s,j}^c) = b_{k,j,r,s}w(X,Y),
\end{equation}
\begin{equation}\label{eq: def v}
v(X,Y,x_{j+r}^w\otimes z_{j+s,j}^c) = b_{k,j,r,s}w(X,Y), \quad
v(X,Y,y_{j+r}^w\otimes z_{j+s,j}^c) = a_{k,j,r,s}w(X,Y).
\end{equation}
where $a_{k,j,r,s}$ and $b_{k,j,r,s}$ will be given later
(satisfying $a_{k,j,r,s}b_{k,j,r,s} = 1$, which makes $u,v,w$ really
a weight scheme). We shall calculate $u_{X,i}$ and $v_{Y,i}$ for $i
= x_{j+r}^w\otimes z_{j+s,j}^c\in set(X)-set(Y)$ ; the other case
$i=y_{j+r}^w\otimes z_{j+s,j}^c$ is just symmetric. Note that if
$x_{j+r}^w\otimes z_{j+s,j}^c\notin set(Y')$ and $X\wedge Y' = k'$,
then $k'\leq j+r$.
\begin{align}
u_{X,x_{j+r}^w\otimes z_{j+s,j}^c} & =
\sum_{k'=1}^{j+r}\sum_{\scriptstyle Y':(X,Y')\in R_P, X\wedge Y'
 =  k'\atop \scriptstyle x_{j+r}^w\otimes z_{j+s,j}^c\notin set(Y')} a_{k',j,r,s}w(X,{Y'}) \\
& \leq  \sum_{k'=1}^{j+r}\sum_{Y':X\wedge Y' = k'}
a_{k',j,r,s}w(X,{Y'}) \\
& =  \sum_{k'=1}^{j+r}a_{k',j,r,s}
\end{align}
The computation for $v_{Y,x_{j+r}^w\otimes z_{j+s,j}^c}$ is a little
more complicated. By definition,
\begin{align}\label{eq: v(y,i)}
  v_{Y,x_{j+r}^w\otimes z_{j+s,j}^c}
& =  \sum_{k'=1}^{j+r}\sum_{\scriptstyle \ X':(X',Y)\in R_P,\
X'\wedge Y = k',\atop \scriptstyle x_{j+r}^w\otimes z_{j+s,j}^c\in
set(X')}
b_{k',j,r,s}w({X'},{Y}) \\
& \leq  \sum_{k'=1}^{j+r}\sum_{\scriptstyle \ X':X'\wedge Y = k',\
\atop \scriptstyle x_{j+r}^w\otimes z_{j+s,j}^c\in set(X')}
b_{k',j,r,s}w({X'},{Y}) \\
\label{eq: passing prob}& =
\sum_{k'=1}^{j+r}b_{k',j,r,s}\pr_{X'}[x_{j+r}^w\otimes
z_{j+s,j}^c\in set(X')|X'\wedge Y = k']
\end{align}
We can see that by adding the clock, the passing probability
$\pr_{X'}[x_{j+r}^w\otimes z_{j+s,j}^c\in set(X')|X'\wedge Y = k']$
is roughly the hitting probability $q(y_{k'-1}^w, y_{k'}^w, k'-1,
x_{j+r}^w, j) + q(y_{k'-1}^w, y_{k'}^w, k'-1, x_{j+r}^w, j+1)$
except for some corner cases. To be more precise, define
\begin{equation}
Bound_{k',j,r,s} = 2p_{j-k'+2}\cdot \lambda[s=1 \text{ OR } j<T] +
2p_{j-k'+1}\cdot \lambda[s=0 \text{ AND } (k'\leq j \text{ OR }
r=0)]
\end{equation}
where the Boolean function $\lambda[\phi] = 1$ if
$\phi$ is true and 0 otherwise. Then

\begin{Claim}\label{clm: Bound}
\quad $\pr_{X'}[x_{j+r}^w\otimes z_{j+s,j}^c\in set(X')|X'\wedge Y =
k'] \leq Bound_{k',j,r,s}$.
\end{Claim}
\begin{proof} We study the probability $\pr_{X'}[x_{j+r}^w\otimes
z_{j+s,j}^c\in set(X')|X'\wedge Y = k']$ case by case. If $s = 1$,
then $r = 1$, and $x_{j+1}^w\otimes z_{j+1,j}^c \in set(X')$ if and
only if $x_{j+1}^w = (x')_{j+1}^w$. So
\begin{equation}\pr_{X'}[x_{j+r}^w\otimes z_{j+s,j}^c\in set(X')|X'\wedge Y =
k'] = q(y_{k'-1}^w, y_{k'}^w, k'-1, x_{j+1}^w, j+1)\leq 2p_{j-k'+2}
\end{equation}
by Lemma \ref{lem: two prob}. If $s = 0$, then $x_{j+r}^w\otimes
z_{j,j}^c \in set(X')$ if and only if ``$x_{j+r}^w = (x')_{j}^w$ or
$x_{j+r}^w = (x')_{j+1}^w$". Also note that
\begin{equation}
\pr_{X'}[x_{j+r}^w = (x')_{j}^w |X'\wedge Y = k'] = q(y_{k'-1}^w,
y_{k'}^w, k'-1, x_{j+r}^w, j)
\end{equation}
unless $k'=j+1$ and $r=1$, in which case $\pr_{X'}[x_{j+r}^w =
(x')_{j}^w |X'\wedge Y = k'] = 0$ because $x_{j+1}^w \otimes
z_{j,j}^c \notin set(Y)$ but $(x')_j^w \otimes z_{j,j}^c = y_j^w
\otimes z_{j,j}^c \in set(Y)$. The other probability
\begin{equation}
\pr_{X'}[x_{j+r}^w = (x')_{j+1}^w |X'\wedge Y = k'] =
\begin{cases} q(y_{k'-1}^w, y_{k'}^w, k'-1, x_{j+r}^w, j+1) & \text{ if } j\leq
T-1 \\
0 & \text{ if } j=T
\end{cases}.
\end{equation} Putting all cases together, we get the desired
result.
\end{proof}

\noindent (Continue the proof of Theorem \ref{thm: general lb}) The
claim implies that
\begin{equation}
v_{Y,x_{j+r}^w\otimes z_{j+s,j}^c} \leq
\sum_{k'=1}^{j+r}b_{k',j,r,s}Bound_{k',j,r,s}.
\end{equation}
The
symmetric case of $u(X,Y,i)$ and $v(X,Y,i)$ where $i$ is a point
$y_{j+r}^w\otimes z_{j+s,j}^c \in set(Y)-set(X)$ can be dealt with
in the same way, yielding $u_{X,y_{j+r}^w\otimes z_{j+s,j}^c} \leq
\sum_{k'=1}^{j+r}b_{k',j,r,s} Bound_{k',j,r,s}$ and
$v_{Y,y_{j+r}^w\otimes z_{j+s,j}^c} \leq
\sum_{k'=1}^{j+r}a_{k',j,r,s}.$

By the definition of $Bound_{k',j,r,s}$, it holds for any $(j,r,s)$
that
\begin{equation}
\sum_{k'=1}^{j+r}Bound_{k',j,r,s}\leq 4\sum_{t=1}^T p_t \qquad
\text{and} \qquad  \sum_{k'=1}^{j+r}\sqrt{Bound_{k',j,r,s}}\leq
4\sum_{t=1}^T \sqrt{p_t}.
\end{equation} Now for the randomized lower bound,
$a_{k',j,r,s} = b_{k',j,r,s} = 1$.
\begin{align}
RLS(G) & = \Omega\left(\min_{j,r,s}
\max\left\{\frac{T-2\sum_{t=1}^{T}p_t}{j+r},
\frac{T-2\sum_{t=1}^{T}p_t}{\sum_{k'=1}^{j+r}Bound_{k',j,r,s}}
\right\}\right) = \Omega\left(\frac{T}{\sum_{t=1}^{T} p_t}\right).
\end{align}
For the quantum lower bound, pick $a_{k',j,r,s} =
\sqrt{Bound_{k',j,r,s}}$, and $b_{k',j,r,s} =
1/\sqrt{Bound_{k',j,r,s}}$. Then
\begin{align}
QLS(G) & =
\Omega\left(\min_{j,r,s}\sqrt{\frac{\left(T-2\sum_{t=1}^{T}p_t\right)\left(T-
2\sum_{t=1}^{T}p_t\right)}{\left(\sum_{k'=1}^{j+r}\sqrt{Bound_{k',j,r,s}}\right)
\left(\sum_{k'=1}^{j+r}\sqrt{Bound_{k',j,r,s}}\right)}}\right) =
\Omega\left(\frac{T}{\sum_{t=1}^{T} \sqrt{p_t}}\right)
\end{align}
This completes the proof of Theorem \ref{thm: general lb}.
\end{proof}

%%%%%%%%%%%%%%%%%%%%%%%%%%%%%%%%%%%%%%%%%%%%%%%%%%%%%%%%%%%%%%%%%%%
%%%%%%%%%%%%%%%%%%%%%%%%%%%%%%%%%%%%%%%%%%%%%%%%%%%%%%%%%%%%%%%%%%%
\section{Applications to the two special graphs} In
this section, we will apply Theorem \ref{thm: general lb} to the two
special graphs. Note that in both cases, the probability $p_t$ is
not easy to upper bound. Also note that we need not only to pick the
random walk, but also the way to decomposed the graph.

\subsection{Lower bounds for Local Search on the
Boolean Hypercube}\label{sec: hypercube lb} To apply Theorem
\ref{thm: general lb} to $\B^n$, we decompose the whole graph into
the two parts $\B^m$ and $\B^{n-m}$, where $m$ is to be decided
later (and to be taken different values for randomized and quantum
lower bounds). Pick the random walk $(\B^m, v_0^w, W)$, where $v_0^w
= 0^m\in \B^m$ and $W(x,t) = \{x^{(i)}: i\in \{0,...,m-1\}\}$ for
each vertex $x=x_0...x_{m-1}\in \B^m$ and each $t\in \mathbb{N}$.
Finally, note that the longest self-avoiding path of the graph
$\B^{n-m}$ is a Hamilton path with length $L = 2^{n-m}-1$.

The following bounds on $p_t$ are rather loose for $10 < t \leq m^2$
but sufficient for our purpose. The proof of the lemma uses some
techniques in generating functions and Fourier analysis.
\begin{Lem} \label{lem: hypercube hit prob}
For any $t\in \mathbb{N}$, we have
\begin{equation} p_t =
\begin{cases}
O(m^{-\lceil t/2 \rceil}) & \text{ if } \ t\leq 10 \\
O(m^{-5}) & \text{ if } \ 10 < t \leq m^2 \\
O(2^{-m}) & \text{ if } \ t > m^2 \\
\end{cases}
\end{equation}
\end{Lem}

\begin{proof} Consider that we put $t$ balls randomly
into $m$ bins one by one. The $j$-th ball goes into the $i_j$-th
bin. Denote by $n_i$ the total number of balls in the $i$-th bin. We
write $n_i\equiv b_i$ if $b_i = n_i \text{ mod } 2$. We say that
$(i_1,...,i_t)$ \emph{generates the parity sequence}
$(b_1,...,b_m)$, or simply $(i_1,...,i_t)$ \emph{generates}
$(b_1,...,b_m)$, if $n_i \equiv b_i$ for all $i\in [m]$. For
$b_1...b_m\in\B^m$, denote by $p^{(t)}[b_1,...,b_m]$ the probability
that $n_i \equiv b_i$, $\forall i\in [m]$. Let $p^{(t)} =
\max_{b_1,...,b_m}p^{(t)}[b_1,...,b_m]$. It is easy to see that
$p^{(t)} = p_t$ in Lemma \ref{lem: hypercube hit prob}, so it is
enough to prove the same bounds in Lemma \ref{lem: hypercube hit
prob} for $p^{(t)}$.

We start with several simple observations. First, we assume that $t$
and $\sum_{i=1}^m b_i$ have the same parity, because otherwise the
probability is 0 and the lemma holds trivially. Second, by the
symmetry, any permutation of $b_1,...,b_m$ does not change
$p^{(t)}[(b_1,...,b_m)]$. Third, $p^{(t)}[(b_1,...,b_m)]$ decreases
if we replace two 1's in $b_1,...,b_m$ by two $0$'s. Precisely, if
we have two $b_i$'s being 1, say $b_1 = b_2 = 1$, then
$p^{(t)}[(b_1,...,b_m)] < p^{(t)}[(0,0,b_3,...,b_m)]$. In fact, note
that
\begin{align}
p^{(t)}[(b_1,...,b_m)] & = \frac{1}{m^t}\sum_{\scriptstyle
n_1+...+n_m = t \atop n_i \equiv b_i, i\in
[m]}\frac{t!}{n_1!...n_m!} \\
& = \frac{1}{m^t}\sum_{\scriptstyle n_3+...+n_m\leq t  \atop
n_i\equiv b_i, i=3,...,m}\left(\frac{t!}{(n_1+n_2)!n_3!...n_m!}
\sum_{\scriptstyle n_1+n_2 = t-n_3-...-n_m \atop  n_i\equiv b_i, i =
1,2}\frac{(n_1+n_2)!}{n_1!n_2!}\right)
\end{align}
where as usual, let $0!=1$. If $n_3+...+n_m < t$, then
\begin{equation}
\sum_{\scriptstyle n_1+n_2 = t-n_3-...-n_m\atop n_i\equiv 1, i =
1,2}\frac{(n_1+n_2)!}{n_1!n_2!} = \sum_{\scriptstyle n_1+n_2 =
t-n_3-...-n_m\atop n_i\equiv 0, i = 1,2}\frac{(n_1+n_2)!}{n_1!n_2!}
\end{equation}
If $n_3+...+n_m = t$, then the only possible $(n_1,n_2)$ is $(0,0)$,
so
\begin{equation}
\sum_{\scriptstyle n_1+n_2 = t-n_3-...-n_m \atop n_i\equiv 1, i =
1,2}\frac{(n_1+n_2)!}{n_1!n_2!} = 0, \qquad \sum_{\scriptstyle
n_1+n_2 = t-n_3-...-n_m\atop n_i\equiv 0, i =
1,2}\frac{(n_1+n_2)!}{n_1!n_2!} = 1.
\end{equation}
Thus $p^{(t)}[(1,1,b_3,...,b_m)]< p^{(t)}[(0,0,b_3,...,b_m)]$.

By the observations, it is sufficient to prove the lemma for the
case $p^{(t)}[(0,...,0)]$ if $t$ is even, and for the case
$p^{(t)}[(1,0,...,0)]$ if $t$ is odd. Note that if $t$ is even, then
\begin{equation}
p^{(t)}[(0,...,0)] = \sum_{i=1}^m\pr[i_1 = i]\pr[(i_2,...,i_t)
\text{ generates } (e_i)]
\end{equation}
where $e_i$ is the $m$-long vector with only coordinate $i$ being 1
and all other coordinates being 0.  By the symmetry, $p^{(t-1)}[e_1]
= ... = p^{(t-1)}[e_{m}]$, thus $p^{(t)}[(0,...0)] = p^{(t-1)}[e_1]
= p^{(t-1)}[1,0,...,0]$. Therefore, it is enough to show the lemma
for even $t$.

We now express $p^{(t)}[0,...,0]$ in two ways. One is to prove the
first case ($t\leq 10$) in the lemma, and the other is for the
second case ($10<t\leq m^2$) and the third case ($t>m^2$) in the
lemma.

To avoid confusion, we write the number $m$ of bins explicitly as
subscript: $p_m^{(t)}[b_1,...,b_m]$. We consider which bin(s) the
first two balls is put into.
\begin{align}
p_m^{(t)}[0,...,0] & = \pr[i_1 = i_2]p_m^{(t-2)}[0,...,0] +
\pr[i_1\neq i_2] p_m^{(t-2)}[1,1,0,...,0]\\
& = \frac{1}{m}p_m^{(t-2)}[0,...,0] +
\frac{m-1}{m}p_m^{(t-2)}[1,1,0,...,0]
\end{align}
To compute $p_m^{(t-2)}[1,1,0,...,0]$, we consider to put $(t-2)$
balls in $m$ bins. By the analysis of the third observations above,
we know that
\begin{align}
& p_m^{(t-2)}[0,...,0] - p_m^{(t-2)}[1,1,0,...,0] \\
= & \pr[n_1 = n_2 = 0, n_3 \equiv 0, ..., n_m \equiv
0] \\
= & \pr[n_1 = n_2 = 0] \pr[n_3 \equiv 0, ..., n_m \equiv 0|n_1=n_2=0] \\
= & \left(\frac{m-2}{m}\right)^{t-2}p_{m-2}^{(t-2)}[0,...,0]
\end{align}
Therefore,
\begin{equation}
p_m^{(t)}[0,...,0] = \frac{1}{m}p_m^{(t-2)}[0,...,0] -
\frac{m-1}{m}\left(\frac{m-2}{m}\right)^{t-2}p_{m-2}^{(t-2)}[0,...,0]
\end{equation}

Now using the above recursive formula and the base case
$p_m^{(2)}[0,...,0] = 1/m$, it is easy (but tedious) to prove by
calculations that $p_m^{(t)}[0,...,0] =
((t-1)!!/m^{\frac{t}{2}})(1-o(1))$ for even $t\leq 10$. This proves
the first case in the lemma.

For the rest two cases, we shall use generating function and some
technique inspired by Fourier analysis. Consider the generating
function
\begin{equation}
(x_1+...+x_m)^t = \sum_{n_1+...+n_m =
t}\binom{t}{n_1,...,n_m}x_1^{n_1}...x_m^{n_m}.
\end{equation}
If $x_i\in \{-1,1\}$, then $(x_1+...+x_m)^t = \sum_{n_1+...+n_m =
t}\binom{t}{n_1,...,n_m}(-1)^{|\{i: x_i = -1, n_i\equiv 1\}|}$. We
sum it over all $x_1...x_m\in \{-1,1\}^m$. Note that for those
$(n_1,...,n_m)$ that has some $n_{i_0}\equiv 1$, it holds due to the
cancelation that $\sum_{x_1,...,x_m\in \{-1,1\}} (-1)^{|\{i: x_i =
-1, n_i\equiv 1\}|} = 0$ . On the other hand, if all $n_i$'s are
even, then $\sum_{x_1,...,x_m\in \{-1,1\}} (-1)^{|\{i: x_i = -1,
n_i\equiv 1\}|} = 2^m$. Thus we have
\begin{equation}
\sum_{x_1,...,x_m\in \{-1,1\}}(x_1+...+x_m)^t =
2^m\sum_{\scriptstyle n_1+...+n_m = t\atop n_i\equiv 0, i\in
[m]}\binom{t}{n_1,...,n_m}.
\end{equation}
And therefore,
\begin{align}
p^{(t)}[0,...,0] & = \frac{1}{m^t}\sum_{\scriptstyle n_1+...+n_m =
t\atop n_i\equiv 0, i\in [m]}\binom{t}{n_1,...,n_m}\\
& = \frac{1}{2^m m^t} \sum_{x_1,...,x_m\in
\{-1,1\}}(x_1+...+x_m)^t \\
& = \frac{1}{2^m m^t}\sum_{i=0}^m\binom{m}{i}(m-2i)^t \\
& =
\frac{1}{2^m}\sum_{i=0}^m\binom{m}{i}\left(1-\frac{2i}{m}\right)^t.
\end{align}
Note that $t$ is even, so $p^{(t)}[0,...,0]$ decreases if $t$
increases by 2, and this proves the second case of the lemma with
the help of the first case. And if $t>m^2/2$, then
\begin{equation}
p^{(t)}[0,...,0] \leq \frac{1}{2^m
}\left(2+\left(1-\frac{2}{m}\right)^t\sum_{i=1}^{m-1}\binom{m}{i}\right)
<
 2/2^m+e^{-m} = O(1/2^m)
\end{equation}
This proves the third case of the lemma.
\end{proof}

Now it is very easy to prove Theorem \ref{thm: hypercube} using this
lemma. For the randomized lower bound, let $m = \lfloor (n+\log_2
n)/2 \rfloor$, then $T = \Theta(2^{n/2}/n^{1/2})$ and
$\sum_{t=1}^{T} p_t = O(1/n)$. Thus $RLS(\B^n) =
\Omega(\sqrt{n}2^{n/2})$. For the quantum lower bound, let
$m=\lfloor(2n+\log_2 n)/3 \rfloor$, then $T =
\Theta(2^{n/3}/n^{1/3})$ and $\sum_{t=1}^{T} \sqrt{p_t} =
O(1/\sqrt{n})$. Thus $QLS(\B^n) = \Omega(2^{n/3}n^{1/6})$.

%// other things to do for in this subsection:

%1. simplify the proof of the lemma, get better bound for the middle t

%%%%%%%%%%%%%%%%%%%%%%%%%%%%%%%%%%%%%%%%%%%%%%%%%%%%%%%%%%%%%%%%%%%
\subsection{Lower bounds for Local Search on the constant
dimensional grid} In this section we shall first prove a lower bound
weaker than Theorem \ref{thm: grid lb} in Section \ref{sec: weaker
grid lb}, and then improve it to Theorem \ref{thm: grid lb} in
Section \ref{sec: better grid} and Section \ref{sec: improvement 2}.

\subsubsection{A weaker family of lower bounds}\label{sec: weaker grid lb}
To simplify notations, we let $n = N^{1/d}$. As in Section \ref{sec:
hypercube lb}, we decompose the grid into two parts, $[n]^m$ and
$[n]^{d-m}$. For each vertex $x = x_0...x_{m-1}\in [n]^m$ and each
$i\in \{0,...,m-1\}$, define
\begin{eqnarray}
x^{(i),-} & = & x_0...x_{i-1}\max\{x_i-1,1\}x_{i+1}...x_{m-1}, \\
x^{(i),+} & = & x_0...x_{i-1}\min\{x_i+1,n\}x_{i+1}...x_{m-1}.
\end{eqnarray}
We perform the random walk $([n]^m, v_0^w, W)$ where $v_0^w =
00...0\in [n]^m$ and
\begin{equation}
W(x,t) = \{x^{((t-1) \text{ mod } m), +}, x^{((t-1) \text{ mod } m),
-}\}.
\end{equation}
To analyze the probability $p_t$ in Theorem
\ref{thm: general lb}, we first consider the following simpler
``line walk". Suppose a particle is initially put at point $i\in
\{1, ..., n\}$, and in each step the particle moves either to
$\max\{1, i-1\}$ or to $\min\{n, i+1\}$, each with probability 1/2.
Let $p_{ij}^{(t)}$ denote the probability that the particle starting
from point $i$ stops at point $j$ after exact $t$ steps of the walk.
%Obviously, we have $\max_{i,j}p_{ij}^{(t)} = 1$ if $t=0$.
For $t\geq 1$, the following proposition gives a very good (actually
tight) estimate on $\max_{ij}p_{ij}^{(t)}$.
\begin{Prop}\label{prop: short walk mixing}
For any $t\geq 1$,
\begin{equation}
\max_{i,j}p_{ij}^{(t)} =
\begin{cases}
O(1/\sqrt{t}) & \text{if}\quad t\leq n^2 \\
O(1/n) & \text{if}\quad t > n^2 \\
\end{cases}
\end{equation}
\end{Prop}

Before the formal proof, let us briefly discuss the main difficulty
and the idea to get around it. First note that since we care about
the whole mixing process (\ie before and after mixing), the standard
eigenvalue gap does not immediately apply. Second, if there are not
the two barriers (1 and $n$) then $p_{ij}^{(t)}$ is very easy to
calculate: $p_{ij}^{(t)} = \binom{t}{t/2+(j-i)/2}$ if $j-i$ and $t$
have the same parity, and 0 otherwise. However, since we now have
the two barriers, it is hard to count the number of paths from $i$
to $j$ after exactly $t$ steps. Fortunately, there is a basic
\emph{reflecting rule} as follows.

\emph{reflecting rule}: In the line walk without barrier, the number
of paths from $i>0$ to $j>0$ in exactly $t$ steps touching or
crossing the point $0$ is equal to the number of paths from $-i$ to
$j$ in exactly $t$ steps.

The proof of this rule is very easy. Suppose a random path touches
the point $0$ at $t$ for the first time, then do a reflection of the
first $t$ steps of the path with respect to point $0$. See Figure
\ref{fig: reflecting rule} for an illustration. It is not hard to
see that this gives a 1-1 correspondence between the following two
sets: 1) the set of paths from $i$ to $j$ after exactly $t$ steps
touching or crossing the point $0$, and 2) the set of paths from
$-i$ to $j$.

\begin{figure}[h]
\begin{center}
\epsfig{file=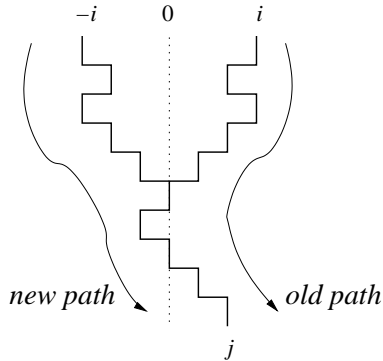, width=5cm} \caption{The proof
of the reflecting rule.} \label{fig: reflecting rule}
\end{center}
\end{figure}

Now let us consider the barrier setting. Note that a path may try to
cross the two barriers in some pattern, for example, try to cross
the left barrier (\ie point 1) for $a$ times and then try to cross
the right barrier (\ie point n) for $b$ times. Imagine that we now
remove the two barriers, then the path will touch (from right) but
not cross the point $1-a$ and will touch (from left) but not cross
the point $n+b-a$. To use the reflecting rule, we just need to
further note the following simple fact:
\begin{equation*}
\begin{array}{rl}
& \{\text{paths touching but not crossing the point } 1-a\} \\
= & \{\text{paths touching or crossing the point } 1-a\}  -
\{\text{paths touching or crossing the point } -a\}.
\end{array}
\end{equation*}
Following this idea, we will construct a series of 1-1
correspondences to reduce the problem step by step to the no-barrier
case. The precise proof is as follows.

\begin{proof}
We consider two settings. One is the line walk on $n$ points $0,
..., n-1$ with the two barriers $0$ and $n-1$ \footnote{Here we let
the $n$ points be $0,...,n-1$ instead of $1,...,n$ just to make the
later calculation cleaner.}. Another is the same except that the
barriers are removed, and we have infinite points in a line. For
each $t$-bit binary string $x=x_1...x_t$, we use $P_i^x$ and $Q_i^x$
to denote the two paths that starting at $i$ and walk according to
$x$ in the two settings. Precisely, at step $s$, $Q_i^x$ goes left
if $x_s = 0$ and goes right if $x_s = 1$ . $P_i^x$ goes in the same
way except that it will stand still if the point is currently at
left (or right) end and it still wants to go left (or right). If the
end point of $P_i^x$ is $j$, then we write $i\rightarrow_t^{P,x} j$.
Let $X_{ij}^{(t),P}$ be the set of $x\in\B^t$ \st
$i\rightarrow_t^{P,x} j$, and put $n_{ij}^{(t),P}=|X_{ij}^{(t),P}|$.
Then by definition, $p_{ij}^{(t)} = n_{ij}^{(t),P}/2^t$. The
notations $i\rightarrow_t^{Q,x} j$, $X_{ij}^{(t),Q}$ and
$n_{ij}^{(t),Q}$ are similarly defined, with the corresponding $P$
changed to $Q$. Note that $n_{ij}^{(t),Q} = \binom{t}{t/2 +
(j-i)/2}$ if $j-i$ and $t$ have the same parity, and 0 otherwise. We
now want to upper bound $n_{ij}^{(t),P}$ in terms of
$n_{ij}^{(t),Q}$.

For a path $P_i^x$, if at some step it is at point $0$ and wants to
go left, we say it \emph{attempts to pass the left barrier}.
Similarly for the right barrier. We say a path is in the $\{a_s,
b_s\}_{s=1}^l$ category if it first attempts to pass the left
barrier for $a_1$ times, and then attempts to pass the right barrier
for $b_1$ times, and so on. We call each round a stage $s$, which
begins at the time that $P_i^x$ attempts to pass the left barrier
for the $(a_1+...+a_{s-1}+1)$-th time, and ends right before the
time that $P_i^x$ attempts to pass the left barrier for the
$(a_1+...+a_{s}+1)$-th time. We also split each stage $s$ into two
halves, cutting at the time right before the path attempts to pass
the right barrier for the $(b_1+...+b_{s-1}+1)$-th time. Note that
$a_1$ may be 0, which means that the path first attempts to pass the
right barrier. Also $b_l$ may be $0$, which means the the last
barrier the path attempts to pass is the left one. But all other
$a_i, b_i$'s are positive. Also note that in the case of $l=0$, the
path never attempts to pass either barrier. Now for any fixed $l>0$,
we consider those categories with $a_1>0$ and $b_l>0$. Other cases
can handled similarly. Partition $X_{ij}^{(t),P}$ as
\begin{equation}
X_{ij}^{(t),P} = \bigcup_{l,\
\{a_s,b_s\}_{s=1}^l}X_{ij}^{(t),P}[\{a_s,b_s\}_{s=1}^l]
\end{equation}
where $X_{ij}^{(t),P}[\{a_s,b_s\}_{s=1}^l]$ contains those $x\in
\B^t$ \st $P_i^x$ is in the category $\{a_s,b_s\}_{s=1}^l$. Put
$n_{ij}^{(t),P}[\{a_s,b_s\}_{s=1}^l] =
|X_{ij}^{(t),P}[\{a_s,b_s\}_{s=1}^l]|$, thus
$n_{ij}^{(t),P}=\sum_{l}\sum_{\{a_s,b_s\}_{s=1}^l}n_{ij}^{(t),P}
[\{a_s,b_s\}_{s=1}^l]$.

Now consider the corresponding paths in $X_{ij}^{(t),Q}$. The
following observation relates $P_i^x$ and $Q_i^x$.

\begin{Obs} For each $x\in X_{ij}^{(t),P}[\{a_s,b_s\}_{s=1}^l]$,
the following two properties hold for any $s$.
\begin{enumerate}
\item In the first half of stage $s$, the path $Q_i^x$ touches
(from right) but does not cross the point $\alpha_s=
\sum_{r=1}^{s-1}(b_r-a_r)-a_s$.

\item In the second half of stage $s$, the path $Q_i^x$ touches
(from left) but does not cross the point $\beta_s =
n-1+\sum_{r=1}^{s}(b_r-a_r)$

\item The path $Q_i^x$ ends at $\gamma =
j+\sum_{s=1}^{l}(b_s-a_s)$
\end{enumerate}
\end{Obs}

We let $Y_{i\gamma}^{(t),Q}[\{\alpha_s,\beta_s\}_{s=1}^l]$ contain
those $x\in \B^t$ satisfying the three conditions in the above
observation, and denote by
$m_{i\gamma}^{(t),Q}[\{\alpha_s,\beta_s\}_{s=1}^l]$ the size of the
set $Y_{i\gamma}^{(t),Q}[\{\alpha_s,\beta_s\}_{s=1}^l]$. Thus the
observation says
$X_{ij}^{(t),P}[\{\alpha_s,\beta_s\}_{s=1}^l]\subseteq
Y_{ij}^{(t),Q}[\{\alpha_s,\beta_s\}_{s=1}^l]$, and therefore we have
$n_{ij}^{(t),P}[\{a_s,b_s\}_{s=1}^l]\leq
m_{i\gamma}^{(t),Q}[\{\alpha_s,\beta_s\}_{s=1}^l]$. So it is enough
to upper bound $m_{i\gamma}^{(t),Q}[\{\alpha_s,\beta_s\}_{s=1}^l]$.

Now for each $x\in
Y_{i\gamma}^{(t),Q}[\{\alpha_s,\beta_s\}_{s=1}^l]$, if we change the
condition 1 in state $s=1$ by allowing the path to cross the point
$\alpha_1$, and let
$Z_{i\gamma}^{(t),Q}[\{\alpha_s,\beta_s\}_{s=1}^l]$ be the new set
satisfying the new conditions, then
$m_{i\gamma}^{(t),Q}[\{\alpha_s,\beta_s\}_{s=1}^l] =
|Z_{i\gamma}^{(t),Q}[\{\alpha_s,\beta_s\}_{s=1}^l]| -
|Z_{i\gamma}^{(t),Q}[\alpha_1-1, \beta_1,
\{\alpha_s,\beta_s\}_{s=2}^l]|$. In other words, the set of paths
touches (from right) but does not cross $\alpha_1$ is the set of
paths touches or crosses $\alpha_1$ minus the set of paths touches
or crosses $\alpha_1-1$.

Now we calculate
$|Z_{i\gamma}^{(t),Q}[\{\alpha_s,\beta_s\}_{s=1}^l]|$ by the
so-called reflection rule. Suppose the first time that $Q_i^x$
touches $\alpha_1$ is $t_{1}$. We reflect the first $t_1$ part of
the path $Q_i^x$ with respect to the point $\alpha_1$. Precisely,
let $y = (1-x_1)...(1-x_{t_{1}})x_{t_{1}+1}...x_t$, then the paths
$Q_i^x$ and $Q_{2\alpha_1-i}^y$ merge at time $t_{1}$. And it is
easy to check that it is a 1-1 correspondence between
$Z_{i\gamma}^{(t),Q}[\{\alpha_s,\beta_s\}_{s=1}^l]$ and
$Y_{2\alpha_1-i,\gamma}^{(t),Q}[\beta_1,\{\alpha_s,\beta_s\}_{s=2}^l]$,
Here
$Y_{2\alpha_1-i,\gamma}^{(t),Q}[\beta_1,\{\alpha_s,\beta_s\}_{s=2}^l]$
is the set of paths starting at $2\alpha_1-i$, satisfying (a) the
condition 2 at the first stage, (b) both conditions 1 and 2 at the
rest $l-1$ stages, and (c) condition 3. So
\begin{align}
|Z_{i\gamma}^{(t),Q}[\{\alpha_s,\beta_s\}_{s=1}^l]| & =
|Y_{2\alpha_1-i,\gamma}^{(t),Q}[\beta_1,\{\alpha_s,\beta_s\}_{s=2}^l]|
=
m_{2\alpha_1-i,\gamma}^{(t),Q}[\beta_1,\{\alpha_s,\beta_s\}_{s=2}^l]
\\ \label{eq: alpha}
& = m_{-2a_1-i,\gamma}^{(t),Q}[\beta_1,\{\alpha_s,\beta_s\}_{s=2}^l]
\\ \label{eq: move}
& =
m_{-a_1-i,\gamma+a_1}^{(t),Q}[\beta_1+a_1,\{\alpha_s+a_1,\beta_s+a_1\}_{s=2}^l]
\end{align}
where \eqref{eq: alpha} is due to the fact that $\alpha_1 = -a_1$,
and \eqref{eq: move} is because that the number of the paths does
not change if we move all the paths right by $a_1$. Similarly, we
have
\begin{align}
|Z_{i\gamma}^{(t),Q}[\alpha_1-1, \beta_1,
\{\alpha_s,\beta_s\}_{s=2}^l]| & =
m_{2\alpha_1-2-i,\gamma}^{(t),Q}[\beta_1,\{\alpha_s,\beta_s\}_{s=2}^l] \\
& =
m_{-a_1-2-i,\gamma+a_1}^{(t),Q}[\beta_1+a_1,\{\alpha_s+a_1,\beta_s+a_1\}_{s=2}^l]
\end{align}
Therefore,
\begin{align}
n_{ij}^{(t),P} [\{a_s,b_s\}_{s=1}^l] & \leq
m_{i\gamma}^{(t),Q}[\{\alpha_s,\beta_s\}_{s=1}^l] \\
& = m_{-2a_1-i,\gamma}^{(t),Q}[\beta_1,\{\alpha_s,\beta_s\}_{s=2}^l]
- m_{-2a_1-2-i,\gamma}^{(t),Q}[\beta_1,\{\alpha_s,\beta_s\}_{s=2}^l]
\\
& =
m_{-a_1-i,\gamma+a_1}^{(t),Q}[\beta_1+a_1,\{\alpha_s+a_1,\beta_s+a_1\}_{s=2}^l]
\\ & \quad -
m_{-a_1-2-i,\gamma+a_1}^{(t),Q}[\beta_1+a_1,\{\alpha_s+a_1,\beta_s+a_1\}_{s=2}^l]
\end{align}
Note that $\alpha_s+a_1 = b_1+\sum_{r=2}^{s-1} (b_r-a_r)-a_s$, \
$\beta_s+a_1 = n-1+b_1+\sum_{r=2}^s(b_r-a_r)$ and $\gamma+a_1 =
j+b_1+\sum_{r=2}^s(b_r-a_r)$ are all functions of $(b_1, a_2,b_2,
..., a_l, b_l)$, not of $a_1$ any more. Therefore,
\begin{align}
& \sum_{a_1,b_1,...,a_l,b_l>0} n_{ij}^{(t),P}[\{a_s,b_s\}_{s=1}^l] \\
\leq & \sum_{b_1,...,a_l,b_l>0}\sum_{a_1>0}
(m_{-a_1-i,\gamma+a_1}^{(t),Q}[\beta_1+a_1,\{\alpha_s+a_1,\beta_s+a_1\}_{s=2}^l]
\\ & \qquad \qquad \qquad \qquad -
m_{-a_1-2-i,\gamma+a_1}^{(t),Q}[\beta_1+a_1,\{\alpha_s+a_1,\beta_s+a_1\}_{s=2}^l])
\\
= & \sum_{b_1,...,a_l,b_l>0}
(m_{-1-i,\gamma+a_1}^{(t),Q}[\beta_1+a_1,\{\alpha_s+a_1,\beta_s+a_1\}_{s=2}^l]
\\
& \qquad \qquad \qquad
+m_{-2-i,\gamma+a_1}^{(t),Q}[\beta_1+a_1,\{\alpha_s+a_1,\beta_s+a_1\}_{s=2}^l])
\\
\leq & \sum_{b_1,...,a_l,b_l>0}
2\max_{h=1,2}\{m_{-h-i,\gamma+a_1}^{(t),Q}[\beta_1+a_1,\{\alpha_s+a_1,\beta_s+a_1\}_{s=2}^l]\}
\end{align}
%Note that due to the parity, only one of
%$m_{-1-i,\gamma+a_1}^{(t),Q}[\beta_1+a_1,\{\alpha_s+a_1,\beta_s+a_1\}_{s=2}^l]$
%and $m_{-2-i,\gamma+a_1}^{(t),Q}[\beta_1+a_1,\{\alpha_s+a_1,\beta_s+a_1\}_{s=2}^l]$
%is nonzero. So the summation of them two items is equal to the
%maximum of them.
Now using the similar methods, \ie reflecting with
respect to points $(n-1+b_1)$ and $(n+b_1)$, moving the paths left
by $b_1$, and finally collapsing the telescope, we can get
\begin{align}
& \sum_{b_1,...,a_l,b_l>0}
m_{-h-i,\gamma+a_1}^{(t),Q}[\beta_1+a_1,\{\alpha_s+a_1,\beta_s+a_1\}_{s=2}^l]
\\
\leq & \sum_{a_2,b_2,...,a_l,b_l>0}
2\max_{k=1,2}\{m_{2n+i+h-k+1,\gamma+a_1-b_1}^{(t),Q}[\{\alpha_s+a_1-b_1,\beta_s+a_1-b_1\}_{s=2}^l]\}
\end{align}
and thus
\begin{align}
& \sum_{a_1,b_1,...,a_l,b_l>0} n_{ij}^{(t),P}[\{a_s,b_s\}_{s=1}^l] \\
\leq & \sum_{a_2,b_2,...,a_l,b_l>0}
4\max_{h=0,1,2}\{m_{2n+i+h,\gamma+a_1-b_1}^{(t),Q}[\{\alpha_s+a_1-b_1,\beta_s+a_1-b_1\}_{s=2}^l]\}
\end{align}
We continue this process, and finally we get
\begin{align}
\sum_{a_1,b_1,...,a_l,b_l>0} n_{ij}^{(t),P}[\{a_s,b_s\}_{s=1}^l] &
\leq
2^{2l}\max_{h=0,1,...,2l}n_{2ln+i+h,\gamma+\sum_{s=1}^l(a_s-b_s)}^{(t),Q} \\
& = 2^{2l} \max_{h=0,1,...,2l}n_{2ln+i+h,j}^{(t),Q}\\
& = 2^{2l} n_{2ln+i,j}^{(t),Q} \\
& \leq 2^{2l} \binom{t}{\frac{t}{2}+\frac{j-i-2ln}{2}}
\end{align}
Thus
\begin{align}
& \sum_{l>0}\sum_{a_1,b_1,...,a_l,b_l>0}
n_{ij}^{(t),P}[\{a_s,b_s\}_{s=1}^l] \\
\leq & \sum_{l\geq 0} 2^{2(l+1)}\binom{t}{\frac{t}{2}+ln} \\
= & 4\binom{t}{t/2} + \sum_{l\geq 1} 2^{2(l+1)}\binom{t}{t/2+ln} \\
\leq & 4\binom{t}{t/2} + \frac{1}{n}\sum_{l\geq 1} 2^{2(l+1)}\left(\binom{t}{t/2+ln}+ \binom{t}{t/2+ln-1} + ... +\binom{t}{t/2+ln-n+1}\right) \\
\leq & 4\binom{t}{t/2} + \frac{1}{n}\sum_{l\geq 1} 2^{2(l+1)}\left(\binom{t}{t}+ \binom{t}{t-1} + ... +\binom{t}{t/2+ln-n+1}\right) \\
\leq & 4\binom{t}{t/2} + \frac{1}{n}\sum_{l\geq 1}
2^{2(l+1)}2^te^{-\frac{2(l-1)^2n^2}{3t}}
\end{align}
where $\binom{t}{t'} = 0$ if $t'
> t$. Here the first two inequalities are by the monotonicity of binomial
coefficients, and the last inequality is by Chernoff's Bound. Now if
$t \leq n^2$, then $\sum_{l\geq 1}
2^{2(l+1)}e^{-\frac{2(l-1)^2n^2}{3t}} \leq \sum_{l\geq 1}
2^{2(l+1)}e^{-\frac{2(l-1)^2}{3}} = O(1)$, so $
\sum_{l>0}\sum_{a_1,b_1,...,a_l,b_l>0}
n_{ij}^{(t),P}[\{a_s,b_s\}_{s=1}^l] \leq O(\binom{t}{t/2}+ 2^t/n) =
O(2^t/\sqrt{t})$. For other categories of $a_1 = 0$ or $b_l = 0$,
the same result can be proved similarly, and the $l=0$ is easy since
$n_{ij}^{(t),Q} = O(2^t/\sqrt{t})$. Putting all things together, we
see that $p_{ij}^{(t)} = O(1/\sqrt{t})$ if $t\leq n^2$. The other
part, \ie $p_{ij}^{(t)} = O(1/n)$ when $t > n^2$, can be easily
derived from this and the fact that $\max_{ij} p_{ij}^{(t)}$
decreases as $t$ increases. This completes our proof.
\end{proof}

Now we use Proposition \ref{prop: short walk mixing} to prove the
weaker lower bounds for grids. Since the random walk $([n]^m, v_0^w,
W)$ is just a product of $m$ line walks, it is not hard to see that
the $p_t$ in the random walk $([n]^m, v_0^w, W)$ satisfies

\begin{equation}
p_t =
\begin{cases}
O(1/\sqrt{t^m}) & \text{ if } t\leq n^2,\\
O(1/n^m) & \text{ if } t > n^2.
\end{cases}
\end{equation}
Now for the randomized lower bounds, when $d>4$ we pick $m=\lceil
d/2 \rceil> 2$ and we get
\begin{equation}
RLS([n]^d) = \Omega\left(\frac{n^{d-m}}{O(1)+n^{d-m}/n^m}\right) =
\Omega(n^{\lfloor d/2 \rfloor}) =
\begin{cases}
\Omega(n^{\frac{d}{2}}) & \text{ if $d$ is odd}, \\
\Omega(n^{\frac{d}{2}-\frac{1}{2}}) & \text{ if $d$ is even}.
\end{cases}
\end{equation}
For $d = 4, 3, 2$, we let $m=2,2,1$ respectively, and get $
RLS([n]^4) = \Omega(n^2/(\log n+1)) = \Omega(n^{2}/\log n)$, $
RLS([n]^3) = \Omega(n/(\log n+1/n)) = \Omega(n/\log n)$, and
$RLS([n]^2) = \Omega(n/(\sqrt{n}+1)) = \Omega(\sqrt{n})$.

For the quantum lower bounds, if $d>6$, we let $m$ be the integer
closest to $2d/3$, thus $m>4$. We get
\begin{equation}
QLS([n]^d) = \Omega\left(\frac{n^{d-m}}{O(1)+n^{d-m}/n^{m/2}}\right)
=
\begin{cases}
\Omega(N^{\frac{1}{3}}) & \text{ if } d = 3d' \\
\Omega(N^{\frac{1}{3}-\frac{1}{3d}}) & \text{ if } d = 3d'+1 \\
\Omega(N^{\frac{1}{3}-\frac{1}{6d}}) & \text{ if } d = 3d'+2 \\
\end{cases}.
\end{equation}
For $d=6$, let $m=4$ and we have $QLS([n]^6) = \Omega(n^2/\log n )$.
For $d = 5, 4, 3$, we let $m=d-2$ and then $QLS([n]^d) =
\Omega(n^2/(n^{2-(d-2)/2}+n^{2-(d-2)/2})) = \Omega(n^{d/2-1})$,
which is $\Omega(n^{5/2}), \Omega(n^{2}), \Omega(n^{3/2})$,
respectively. For $d=2$, let $m=1$ and $QLS([n]^2) =
\Omega(\frac{n}{n^{3/4}}) = \Omega(n^{1/4})$.

%// other things to do:

%1. prove the proposition is tight

%2. eigenvalue gap: what's known results? Are they enough to prove
%our proposition? Will our proposition improve known results?

%%%%%%%%%%%%%%%%%%%%%%%%%%%%%%%%%%%%%%%%%%%%%%%%%%%%%%%%%%%%%%%%%%%
\subsubsection{Improvement}\label{sec: better grid} One weakness of
the above proof is the integer constraint of the dimension $m$. We
now show a way to get around the problem, allowing $m$ to be any
real number between 0 and $d-1$. The idea is to partition the grid
into many blocks, with different blocks representing different time
slots, and the blocks are threaded into one very long block by many
paths that are pairwise disjoint. Roughly speaking, we view $[n]^d$
as the product of $d$ line graph $[n]$. For each of the first $d-1$
line graphs, we cut it into $n^{1-r}$ parts evenly, each of size
$n^r$. (Here $r = m/(d-1)$). Then $[n]^{d-1}$ is partitioned into
$n^{(d-1)(1-r)}$ smaller grids, all isomorphic to $[n^r]^{d-1}$.
Putting the last dimension back, we have $n^{(d-1)(1-r)}$
\emph{blocks}, all isomorphic to $[n^r]^{d-1} \times [n]$. Now the
random walk will begin in the first block, and within each block, it
is just one step of random walk in $[n^r]^{d-1}$ followed by two
steps of one-way walk in the last dimension space $[n]$. When the
walk runs out of the clock $[n]$, the walk will move to the next
block via a particular block-changing path. All block-changing paths
are carefully designed to be disjoint, and they ``thread" all the
blocks to form a $[n^r]^{d-1} \times [L]$ grid, where $L =
(n-2n^r)n^{(1-r)(d-1)}$. ($L$ is not $n\cdot n^{(1-r)(d-1)}$ because
we need $2n^r$ points for the block-changing paths.) Figure
\ref{fig: reflect} is an illustration for the case of $d=2$.

We now describe the partition and the walk precisely. For $x =
x_0...x_{d-1}$ in $[n]^d$, let $x^{(k)= l} =
x_0...x_{k-1}lx_{k+1}...x_{d-1}$, and $x^{(k) = (k) + i} =
x_0...x_{k-1}(x_k+i)x_{k+1}...x_{d-1}$, where $i$ satisfies
$x_k+i\in [n]$. Recall that $x^{(i),-} = x^{(i)=\max\{x_i-1,1\}}$
and $x^{(i),+} = x^{(i)=\min\{x_i+1,n\}}$.

\begin{figure}[h]
\begin{center}
\epsfig{file=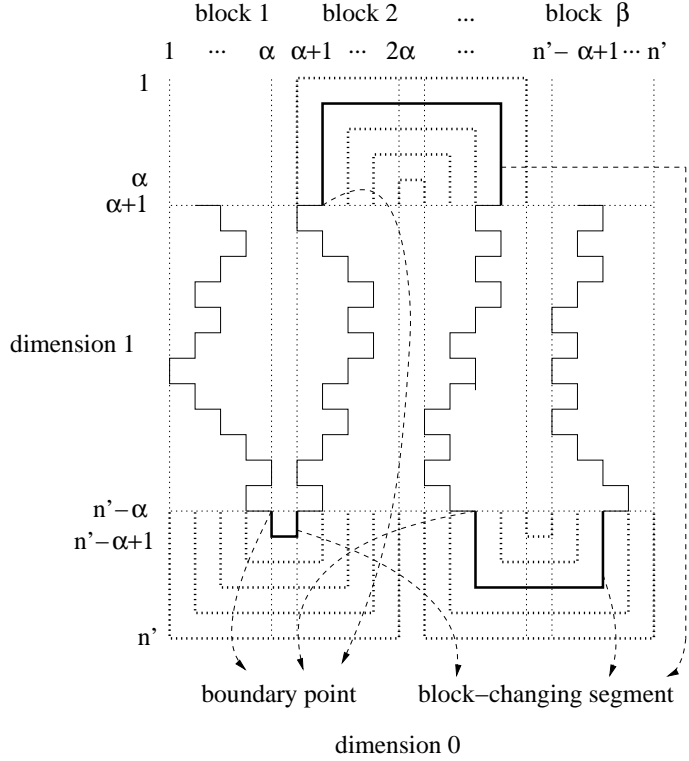, width=9cm} \caption{Illustration for
changing a block in the 2-dimensional grid} \label{fig: reflect}
\end{center}
\end{figure}

For any fixed constant $r\in (0,1)$, let $\alpha = \lfloor n^r
\rfloor$, $\beta = \lfloor n^{1-r} \rfloor$ and $n' = \alpha\beta$.
Note that $n' \geq (n^r-1)(n^{1-r}-1) = n - o(n)$. We now consider
the slightly smaller grid $[n']^d$. Let $V_1$ be the set $[n']^{d-1}
= \{x_0...x_{d-2}: x_i\in [n']\}$. We cut $V_1$ into $\beta^{d-1}$
parts $\{x_0...x_{d-2}: (k_i-1)\alpha < x_i \leq
k_i\alpha\}_{k_0...k_{d-2} \in [\beta]^{d-1}}$, each of which is a
small grid isomorphic to $[\alpha]^{d-1}$. We then refer to the set
$\{x_0...x_{d-2}x_{d-1}: (k_i-1)\alpha < x_i \leq k_i\alpha, i =
0,...,d-2, \alpha < x_{d-1}\leq n'-\alpha\}$ as the ``\emph{block}
$(k_0,...,k_{d-2})$". Note that $(k_0,...,k_{d-2})$ can be also
viewed as a point in grid $[\beta]^{d-1}$, and there is a Hamilton
path $HamPath_{\beta,d-1}$ in $[\beta]^{d-1}$, as defined in Section
2. We call the block $(k'_0,...,k'_{d-2})$ \emph{the next block} of
the block $(k_0, ..., k_{d-2})$ if $(k'_0,...,k'_{d-2})$, viewed as
the point in $[\beta]^{d-1}$, is the next point of $(k_0, ...,
k_{d-2})$ in $HamPath_{\beta,d-1}$. Note that by our definition of
$HamPath_{\beta,d-1}$, we know that $\exists i\in \{0,...,d-2\}$ \st
$k'_{i} \in \{k_{i}+1, k_i-1\}$ and for all other $j\neq i$, $k'_j =
k_j$. That is, adjacent blocks have only one coordinate to be
different, and this difference is 1. We call the the block $(k_0,
..., k_{d-2})$ \emph{the last block} if $(k_0, ..., k_{d-2})$ is the
last point in $HamPath_{\beta,d-1}$.

Now we define the random walk by describing how a particle may go
from start to end. The path set is just all the possible paths the
particle goes along. Intuitively, within one block, the last
dimension $d-1$ serves as the clock space. So as before, we perform
one step of line walk (in the dimension which is the circularly next
dimension of the last one that the walk just goes in), followed by
two steps of walk in the clock space. If we run out of clock, we say
we reach \emph{a boundary point} at the current block, and we move
to the next block via a path segment called \emph{block-changing
segment}. In what follows, we specify how the particle may move
during the whole random walk process, including going through
block-changing segments. We always use $x_0...x_{d-1}$ to denote the
current position of the particle, and assume $x_i = (k_i - 1)\alpha
+ y_i$, \ie $x$ is in the block $(k_0,...,k_{d-2})$ with the offsets
$(y_0, ..., y_{d-1})$. Thus the instruction $x_0 = x_0 + 1$, for
example, means that the particle moves from $x_0...x_{d-1}$ to
$(x_0+1)x_1...x_{d-1}$.

\begin{enumerate}
\item Initially  $x_0 = ... = x_{d-2} = 0$,
$x_{d-1} = \alpha+1$, $k_0 = ... = k_{d-2} = 1$.

\item \textbf{for} $t = 1$ \textbf{to}
$(n'-2\alpha)\beta^{d-1}$,

\quad Let $t' = \lfloor \frac{t-1}{n'-2\alpha} \rfloor$, $i=(t-1)
\text{ mod } (d-1)$

\quad \textbf{do} either $x_{i} = \max\{x_{i} - 1,
(k_i-1)\alpha+1\}$ or $x_{i} = \min\{x_{i} + 1, k_i\alpha\}$
randomly

\quad \textbf{if} $t \neq k(n'-2\alpha)$ for some positive integer
$k$,

\quad \quad \textbf{do} $x_{d-1} = x_{d-1}+(-1)^{t'}$ twice

\quad \textbf{else} \quad (\emph{the particle is now at a boundary
point})

\quad \quad \textbf{if} the particle is not in the last block \quad

\quad \quad (Suppose the current block changes to the next block by
increasing $k_j$ by $b\in \{-1, 1\}$)

\quad \quad \quad \textbf{do} $x_{d-1} = x_{d-1}+(-1)^{t'}$
\textbf{for} $(\alpha+1-y_j)$ times

\quad \quad \quad \textbf{do} $x_{j} = x_{j}+b$ \textbf{for}
$2(\alpha+1-y_j)-1$ times

\quad \quad \quad \textbf{do} $x_{d-1} = x_{d-1}+(-1)^{t'+1}$
\textbf{for} $(\alpha+1-y_j)$ times

\quad \quad \quad $k_j = k_j + b$

\quad \quad \textbf{else}

\quad \quad \quad The particle stops and the random walk ends

\end{enumerate}

It is easy to verify that every boundary point has one unique
block-changing segment, and different block-changing segments do not
intersect. Also note that we do not let the clock tick when we are
moving from one block to the another. Thus the block-changing
segments thread all the blocks to form a $[\alpha]^{d-1}\times [L]$
grid, where $L = (n'-2\alpha)\beta^{d-1}$. Actually, for our lower
bound purpose, we can think of the random walk as performed in the
product graph $[\alpha]^{d-1}\times [L]$. We will next make this
clearer as below.

What we care about is, as before, the probability that the random
walk starting from a point $x = x_0...x_{d-1}$ passes another point
$x' = x'_0...x'_{d-1}$. Note that for any point $x$ (including those
on the block-changing segments), there is only one time $t$ when the
walk may hit $x$, and this $t$ is known by $x$ itself. Similarly we
use $t'$ to denote the time when the path passes $x'$. Denote the
probability that the random walk starting from $x$ passes $x'$ by
$\pr[x\rightarrow x']$. As before suppose $x_i = (k_i-1)\alpha+y_i$
and $x'_i = (k'_i-1)\alpha+y'_i$ for $i\in \{0,...,d-2\}$.

We first consider the case that one of the two points, say $x'$ is
on a block-changing segment. Since different block-changing segments
never intersect, a path passes $x'$ if and only if the path passes
the boundary point $x''$ at the beginning of the block-changing
segment that $x'$ is in. Also note that the time that the path
passes $x''$ is also $t'$ because the time does not elapse on the
block-changing segment. So it holds that $\pr[x\rightarrow x'] =
\pr[x\rightarrow x'']$, and it is enough to consider the case that
both $x$ and $x'$ are not in clock-changing segments.

Now suppose both $x$ and $x'$ are not in clock-changing segments. In
general, $x$ and $x'$ may be not in the same block , so going from
$x$ to $x'$ needs to change blocks. Recall that to change from the
block $(k_0,...,k_{d-2})$ to the next one, only one $k_i$ changes by
increasing or decreasing by 1. Suppose that to go to $x'$ from $x$,
we change blocks for $c$ times, by changing $k_{i_1}, k_{i_2}, ...,
k_{i_c}$ in turn. Let $n_j = |\{s\in [c]: i_s = j\}|$. Note that to
get to $x'$ from $x$ after $t'-t$ steps, the coordinate $j$ needs to
be $x'_j$ after $t'-t$ steps for each coordinate $j\in
\{0,...,d-2\}$. It is not hard to see that if a block-changing needs
to change $k_j$ by increasing $b\in \{-1, 1\}$, then among all the
offsets $y_i$'s,  only the $y_j$ gets changed, and the change is a
reflection within the block. That is, suppose $x_j$ is
$(k_j-1)\alpha + y_j$ before the block-changing, then $x_j$ changes
to $(k_j+b-1)\alpha + (\alpha+1-y_j)$ after the block-changing. So
if $c=1$, then $\pr[x\rightarrow x']$ is equal to the probability
that a random walk in $[\alpha]^{d-1}$ starting from $y_0...y_{d-2}$
hits $y''_0...y''_{d-2}$ after exactly $t'-t$ steps, where $y''_j =
y'_j$ if $j\neq i_1$ and $y''_{i_1} = \alpha+1-y'_{i_1}$. For
general $c$, $\pr[x\rightarrow x']$ is equal to the probability that
a random walk in $[\alpha]^{d-1}$ starting from $y_0...y_{d-2}$ hits
$y''_0...y''_{d-2}$ after exactly $t'-t$ steps, where $y''_j = y'_j$
if $n_j$ is even and $y''_{j} = \alpha+1-y'_{j}$ if $n_j$ is odd.
Note that this probability has nothing to do with the
block-changing; it is just the same as we have a clock space
$[(n'-2\alpha)\beta^{d-1}]$ to record the random walk on
$[\alpha]^{d-1}$. Thus we can use Proposition \ref{prop: short walk
mixing} to upper bound this probability and just think of the graph
as $[n^r]^{d-1}\times [L]$ and use Theorem \ref{thm: general lb},
with $G^w = [n^r]^{d-1}$ and $G^c = [L]$.

Now we have $T = \lfloor L/2 \rfloor$ and $p_t =
O(1/\sqrt{t^{d-1}})$ for $t\leq n^{2r}$ and $p_t = O(1/n^{r(d-1)})$
for $t> n^{2r}$. So for randomized lower bounds, if $d\geq 4$, then
let $r = d/(2d-2)$ and we get
\begin{equation}
RLS([n]^d) =
\Omega\left(n^{1+(1-r)(d-1)}/\left(\sum_{t=1}^{n^{d/(d-1)}}\frac{1}{\sqrt{t^{d-1}}}
+ \frac{n^{1+(1-r)(d-1)}}{n^{r(d-1)}}\right)\right) =
\Omega\left(n^{d/2}\right).
\end{equation}
If $d=3$, let $r=3/4 - \log\log n / (4\log n)$, and we get
$RLS([n]^3) = \Omega((n^3/\log n)^{1/2})$. For $d=2$, let $r = 2/3$
and we get $RLS([n]^2) = \Omega(n^{2/3})$.

For the quantum lower bounds, if $d\geq 6$, then let $r = 2d/(3d-3)$
and we get
\begin{equation}
QLS([n]^d) =
\Omega\left(n^{1+(1-r)(d-1)}/\left(\sum_{t=1}^{n^{d/(d-1)}}\frac{1}{t^{(d-1)/4}}
+ \frac{n^{1+(1-r)(d-1)}}{n^{r(d-1)/2}}\right)\right) =
\Omega(n^{d/3}).
\end{equation}
If $d=5$, then let $r = 5/6-\log \log n / (6 \log n)$ and
$QLS([n]^5) = \Omega((n^5/\log n)^{1/3})$. For $2\leq d\leq 4$, we
let $r = d/(d+1)$, then $QLS([n]^d) = \Omega(n^{d/2-d/(d+1)})$,
which is $\Omega(n^{1/3})$, $\Omega(n^{3/4})$, $\Omega(n^{6/5})$ for
$d = 2, 3, 4$, respectively.

%%%%%%%%%%%%%%%%%%%%%%%%%%%%%%%%%%%%%%%%%%%%%%%%%%%%%%%%%%%%%%%%%%%
\subsubsection{Further improvement on 2-dimensional grid $[n]^2$}\label{sec:
improvement 2}

Some other random walk may be used to further improve the lower
bound on low dimension grid cases. Here is one way to improve
$QLS([n]^2)$ from $\Omega(n^{1/3})$ to $\Omega(n^{4/5})$. We cut the
graph $[n]^2$ into $n^{2/5}$ smaller grids, each of size
$n^{4/5}\times n^{4/5}$. Without loss of generality, assume both
$n^{1/5}$ and $n^{4/5}$ are integers, and further assume $n^{1/5} =
3$ mod 4; otherwise we can consider a slightly smaller grid by the
simple trick as at the beginning of Section \ref{sec: better grid}.
We shall use a random walk similar to Aaronson's in \cite{Aa04} as
follows in each block, and change blocks after each step. Thus
different blocks to record different time.

For any time $t\in [n^{1/5}(n^{1/5}-1)]$, suppose $t = 2rn^{1/5} +
t'$ where $r\in \{0,1, ..., (n^{1/5}-3)/2\}$ and $t' \in \{1,2,...,
2n^{1/5}\}$. Let
\begin{equation}
u =
\begin{cases}
0 & \text{ if } t' \equiv 0,1 \text{ (mod 4)}\\
n^{4/5} & \text{ if } t' \equiv 2,3 \text{ (mod 4)}
\end{cases}
\end{equation}
Let $block(t)$ as the small grid
\begin{equation}
\begin{cases}
\{(x = (\lceil t'/2 \rceil - 1)n^{4/5} + x', y = 2rn^{4/5} + u +
y'): x', y' \in [n^{4/5}] \} & \text{ if } r \text{ is even} \\
\{(x = (n^{1/5} - \lceil t'/2 \rceil )n^{4/5} + x', y = 2rn^{4/5} +
u + y'): x', y' \in [n^{4/5}] \} & \text{ if } r \text{ is odd}
\end{cases}
\end{equation}
The $(x',y')$ is called the offset of $(x,y)$. Now define the random
walk as follows and as depicted in Figure \ref{fig: 2D walk}.

\begin{figure}[h]
\begin{center}
\epsfig{file=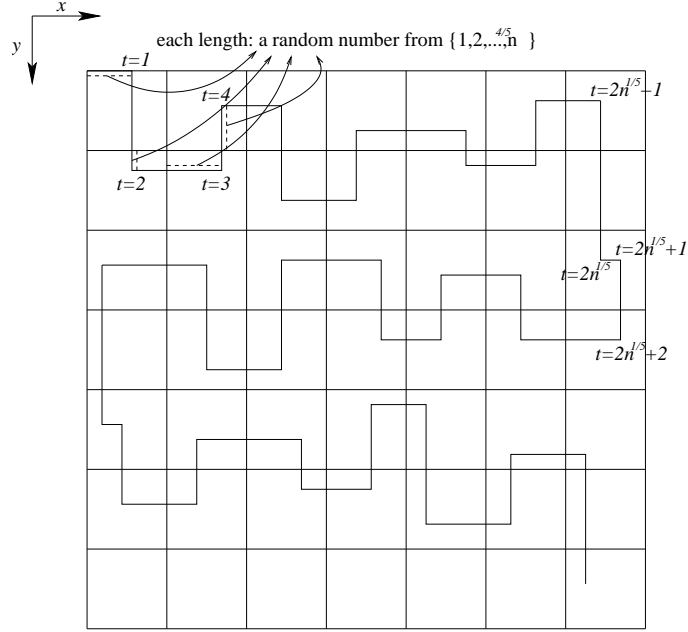, width=9cm}

\caption{A different random walk in the 2 dimensional-grid}

\label{fig: 2D walk}
\end{center}
\end{figure}%\vspace{-1em}

\vspace{1em}

Initially $(x,y) = (1,1)$

\textbf{for} $t = 1, 2, ..., n^{1/5}(n^{1/5}-1)$ \quad (Suppose the
current point is $(x,y)$ with offset $(x',y')$)

\quad \textbf{if} $t'$ is odd

\quad \quad pick a random $x''\in [n^{1/5}]$, move horizontally to
the point in $block(t)$ with the offset $(x'', y')$

\quad \textbf{else}

\quad \quad \textbf{if} $t' = 2n^{1/5}$ \textbf{then} c = 1
\textbf{else} c = 0

\quad \quad pick a random $y''\in [n^{1/5}]$, move vertically to the
point in $block(t+c)$ with the offset $(x', y'')$

\vspace{1em}

We then follow the same track as in the proof of Theorem \ref{thm:
general lb}. To get a reduction from Local Search on $[n]^2$ to the
$Path_P$ problem, we define the function
\begin{equation}
f_X(v) =
\begin{cases}
|v - (1,1)|_{[n]^2} & \text{ if } v\notin set(X) \\
-2n^{4/5}(t-1) - (-1)^r x'_v + (-1)^{\lceil t'/2 \rceil} y'_v &
\text{ if } v\in set(X)\cap block(t)
\end{cases}
\end{equation}
Intuitively, the function value decreases along the path as before.
But the decrement is not always by 1: each block has its fixed value
setting. If for example the path passes through the block toward
right and down (as in the first block), then the value $-x'-y'$ is
used within the block. In this way, we do not need to know the
length of the path segment from top to $v$ to calculate each
$f_X(v)$.

What we care about is still, as in Equality \eqref{eq: passing
prob}, the probability that the path $X'$ passes another point $x$
on $X$, under the condition that $X'\wedge Y = k'$. It is not hard
to see that this probability is $\Theta(1)$ in general if $x$ is in
$block(k')$, and $\Theta(1/n^{4/5})$ otherwise (\ie when $x$ is in
$block(t)$ for some $t> k'$). Thus by $L = \Theta(n^{2/5})$ we have
\begin{equation}
QLS([n]^2) =
\Omega\left(n^{2/5}/\left(1+n^{2/5}/\sqrt{n^{4/5}}\right)\right) =
\Omega(n^{2/5})
\end{equation}
This completes the proof of Theorem \ref{thm: grid lb}.

Note that this random walk suffers from the fact that the ``passing
probability" is now $n^{4/5}$ times the ``hitting probability". So
for general $d$, we can get $RLS([n]^d) = \Omega(n^{d/(d+1)})$ and
$QLS([n]^d) = \Omega(n^{d/(2d+1)})$, which only gives better results
for $QLS$ on the 2-dimensional grid.

%%%%%%%%%%%%%%%%%%%%%%%%%%%%%%%%%%%%%%%%%%%%%%%%%%%%%%%%%%%%%%%%%%%
%%%%%%%%%%%%%%%%%%%%%%%%%%%%%%%%%%%%%%%%%%%%%%%%%%%%%%%%%%%%%%%%%%%
\section{New algorithms for Local Search on general graphs}\label{sec:
upper bound}

In \cite{Al83, Aa04}, a randomized and a quantum algorithm for Local
Search on general graphs are given as follows. Do a random sampling
over all the vertices, find a vertex $v$ in them with the minimum
$f$-value. (For the minimum $f$-value finding procedure, The
randomized algorithm in \cite{Al83} just queries all these vertices
and find the minimum, while the quantum algorithm in \cite{Aa04}
uses the algorithm by Durr and Hoyer \cite{DH96} based on Grover
search \cite{Gr96} to get a quadratic speedup.) If $v$ is a local
minimum, then return $v$; otherwise we follow a \emph{decreasing
path} as follows. Find a neighbor of $v$ with the minimum $f$-value,
and continue this minimum-value-neighbor search process until
getting to a local minimum. We can see that the algorithms actually
fall into the generic algorithm category (see Section 1), with the
initial point picked as the best one over some random samples.

In this section, we give new randomized and quantum algorithms,
which work better than this simple ``random sampling + steepest
descent" method when the graph expands slowly. Here the idea is that
after finding the minimum vertex $v$ of the sampled points, instead
of following the decreasing path of $v$, we start over within a
smaller range, which contains those vertices ``close to" $v$. If
this smaller range contains a local minimum for sure, then we can
simply search a local minimum in it and do this procedure
recursively. But one caveat here is that a straightforward recursion
does not work, because a local minimum $u$ in the smaller range may
be not a local minimum in the original larger graph $G$ (since $u$
may have more neighbors in $G$). So we shall find a small range
which has a ``good" boundary in the sense that all vertices on the
boundary have a large $f$-value.

Now we describe the algorithm precisely, with some notations as
follows. For $G = (V,E)$, a given function $f:V\rightarrow
\mathbb{N}$, a vertex $v\in V$ and a set $S\subseteq V$, let $n(v,S)
= |\{u\in S: f(u)<f(v)\}|$. The boundary $B(S)$ of the set
$S\subseteq V$ is defined by $B(S) = \{u\in S: \exists v\in V-S \
\st \ (u,v)\in E\}$. In particular, $B(V) = \emptyset$. A decreasing
path from a vertex $v\in V$ is a sequence of vertices $v_0,
v_1,...,v_k$ such that $v_0 = v$, $v_k$ is a local minimum and
$f(v_{i+1}) = \min_{v: (v_i,v)\in E}f(v) < f(v_{i})$ for
$i=0,...,k-1$. We write $f(u)\leq f(S)$ if $f(u)\leq f(v)$ for all
$v\in S$. In particular, it always holds that $f(u) \leq
f(\emptyset)$. Suppose $d = \max_{u,v\in V}|u-v|$ is the diameter of
the graph, and $\delta = \max_{v\in V} |\{u: (u,v)\in E\}|$ is the
max degree of the graph. In the following algorithm, the
asymptotical numbers at the end of some command lines are the
numbers of randomized or quantum queries needed for the step. For
those commands without any number, no query is needed.

\begin{enumerate}

\item \label {step: initialize} $m_{0} = d$, $U_{0} = V$;

\item $i = 0$;

\item \textbf{while} ($|m_{i}| > 10$) \textbf{do}

  \begin{enumerate}
  \item \label{step: sample} Randomly pick (with replacement) $\lceil\frac{8|U_{i}|}{m_{i}}\log
  \frac{1}{\epsilon_1}\rceil$
  vertices from $U_{i}$, where $\epsilon_1 = 1/(10\log_2 d)$;

  \item \label{step: min search} Search the sampled vertices for one $v_{i}$ with the minimal
  $f$ value.

  - Randomized algorithm: query all the sampled vertices and get
  $v_{i}$. \qquad --- $O\left(\frac{8|U_{i}|}{m_{i}} \log \frac{1}{\epsilon_1}\right)$

  - Quantum algorithm: use Durr and Hoyer's algorithm \cite{DH96} with the
  error probability at most $\epsilon_2 = 1/(10\log_2 d)$.
  \qquad --- $O\left(\sqrt{\frac{8|U_{i}|}{m_{i}} \log \frac{1}{\epsilon_1}}\log \frac{1}{\epsilon_2}\right)$

  \item \label{step: point update} \textbf{if} $i=0$, \textbf{then} $u_{i+1} = v_{i}$;

  \textbf{else} \textbf{if} $f(u_{i}) \leq f(v_{i})$, \textbf{then} $u_{i+1} =
  u_{i}$;

  \qquad \textbf{else} $u_{i+1} = v_{i}$;

  \item \label{step: boundary search} \textbf{for} $j = 1, 2, ...$
  \begin{enumerate}
  \item \label{step: boundary sample} Randomly pick $m_{ij}\in
  M_{i} =
  \{m: m_i/8\leq m \leq m_i/2,\ |W(m)| \leq 10|U_i|/m_i\}$, where $W(m)= \{w\in U_{i}: |w -
  u_{i+1}| = m\}$. Let $W_{ij} = W(m_{ij})$.

  \item \label{step: boundary test} Test whether $f(u_{i+1}) \leq f(W_{ij})$

  - Randomized algorithm: query all vertices in $W_{ij}$. \qquad --- $O(|W_{ij}|)$

  - Quantum algorithm: use Durr and Hoyer's algorithm \cite{DH96} on $W_{ij}$ with the
  error probability at most $\epsilon_3 = 1/(200\log_2 d)$. \qquad --- $O\left(\sqrt{|W_{ij}|}\log \frac{1}{\epsilon_3}\right)$

  \item If the answer is Yes, jump out of this \textbf{for} loop and go to Step \ref{step: length and range update}.

  \end{enumerate}

  \item \label{step: length and range update} $J_i = j$, $m_{i+1} =
  m_{ij}$, $W_{i} = W_{ij}$,
  $U_{i+1}= \{u\in U_{i}: |u-u_{i+1}|\leq m_{i+1}\}$;

  \item $i = i + 1$;
  \end{enumerate}

\item $I = i$;

\item \label{step: final search} Follow a decreasing path of
$u_{I}$ to find a local minimum.

- Randomized algorithm: in each step, query all the neighbors \qquad
--- $O(\delta)$

- Quantum algorithm: in each step, use Durr and Hoyer's algorithm
with the error probability at most $1/100$ \qquad ---
$O(\sqrt{\delta})$
\end{enumerate}

Define $c(k) = \max_{v\in V} |\{u: |u-v| \leq k\}|$. Apparently, the
expanding speed of a graph is upper bounded by $c(k)$. The following
theorem says that the algorithm is efficient if $c(k)$ is small.

\begin{Thm}\label{thm: upper bound}
The algorithm outputs a local minimum with probability at least 1/2.
The randomized algorithm uses $O\left(\sum_{i=0}^{I-1}
\frac{c(m_{i})}{m_{i}} \log \log d \right)$ queries in expectation,
and the quantum algorithm uses $O\left(\sum_{i=0}^{I-1}
\sqrt{\frac{c(m_{i})}{m_{i}}} (\log \log d)^{1.5} \right)$ queries
in expectation.

In case that $c(k) = O(k^\alpha)$ for some $\alpha \geq 1$ and $k =
1, ..., d$, the expected number of queries that the randomized
algorithm uses is $O\left( \frac{d^{\alpha-1} -1}{1-2^{1-\alpha}}
\log \log d \right)$ if $\alpha > 1$ and $O(\log d \log \log d)$ if
$\alpha = 1$. The expected number of queries that the quantum
algorithm use is $O\left( \frac{d^{\frac{\alpha-1}{2}} -1}{1 -
2^{\frac{1-\alpha}{2}} } (\log \log d)^{1.5} \right)$ if $\alpha >
1$ and $O(\log d \log log d)$ if $\alpha = 1$.
\end{Thm}

Several comments before proving the theorem:
\begin{enumerate}
\item $\lim_{\alpha\rightarrow 1}\frac{d^{\alpha-1}
-1}{1-2^{1-\alpha}} = \lim_{\alpha\rightarrow 1}
\frac{d^{\frac{\alpha-1}{2}} -1}{1 - 2^{\frac{1-\alpha}{2}} } =
\log_2 d$

\item If $\alpha - 1 \geq \epsilon$ for some constant $\epsilon > 0$, then $\frac{d^{\alpha-1}
-1}{1-2^{1-\alpha}} = \Theta(d^{\alpha - 1})$ and
$\frac{d^{\frac{\alpha-1}{2}} -1}{1 - 2^{\frac{1-\alpha}{2}} } =
\Theta(d^{(\alpha - 1)/2})$.

If further the bound $c(k) = O(k^\alpha)$ is tight in the sense that
$N = c(d) = \Theta(d^\alpha)$, then $RLS(G) = O\left(\frac{N}{d}\log
\log d\right)$ and $QLS(G) = O\left(\sqrt{\frac{N}{d}}(\log \log
d)^{1.5}\right)$.

\item For 2-dimensional grid, $d = \Theta(n)$ and $\alpha = 2$. Thus
Theorem \ref{thm: grid ub} follows immediately.

\end{enumerate}

\begin{proof}
We shall prove the theorem for the quantum algorithm. The analysis
of the randomized algorithm is almost the same (and actually
simpler). We say $W_i$ is \emph{good} if $f(u_{i+1}) \leq f(W_i)$.
We shall first prove the following claim, then the theorem follows
easily.

\begin{Claim}
For each $i = 0, 1, ..., I-1$, the following three statements hold.
\begin{enumerate}
\item \label{statement: tail length bound} $n(u_{i+1}, U_{i+1}) \leq
n(u_{i+1}, U_i) \leq m_i/8 \leq m_{i+1}$ with probability
$1-\epsilon_1-\epsilon_2$.

\item \label{statement: good boundary} If $n(u_{i+1}, U_i) \leq m_i/8$, then $W_i$ is good with
probability $1-\epsilon_3J_i$, and $\av[J_i] \leq 2$.

\item \label{statement: all boundary good} If $W_0, ..., W_i$ are all good, then $f(u_{i+1}) \leq
f(B(U_{i+1}))$, and $u_{i+1} \notin B(U_{i+1})$.
\end{enumerate}
\end{Claim}

\begin{proof}
\ref{statement: tail length bound}: In Step \ref{step: sample} -
\ref{step: point update}, denote by $S$ the set of the $\lceil
\frac{8|U_{i}|}{m_{i}}\log \frac{1}{\epsilon_1}\rceil $ sampled
vertices in Step \ref{step: sample}. Let $a = \min_{u\in S} f(u)$,
then $|\{v\in U_{i}: f(v) < a\}| \leq m_{i}/8$ with probability at
least $1-\epsilon_1$. The $v_i$ found in Step \ref{step: min search}
achieves the minimum in the definition of $a$ with probability at
least $1-\epsilon_2$. Put the two things together, we have
$n(v_{i},U_{i})\leq m_{i}/8$ with probability at least
$1-\epsilon_1-\epsilon_2$. Since $f(u_{i+1}) \leq f(v_{i})$ (by Step
\ref{step: point update}), $U_{i+1} \subseteq U_i$ (by Step
\ref{step: length and range update}) and $m_{i+1} \geq m_i/8$ (by
Step \ref{step: boundary sample}), we have $n(u_{i+1}, U_{i+1}) \leq
n(u_{i+1}, U_i) \leq n(v_i, U_i) \leq m_i/8 \leq m_{i+1}$ with
probability at least $1-\epsilon_1-\epsilon_2$.

\ref{statement: good boundary}: We say an $m_{ij}$ is good if the
corresponding $W_{ij}$ is good, \ie $f(u_{i+1}) \leq f(W_{ij})$.
Note that for any $m_{ij}\in [m_{i}]$, we have $W_{ij}\subseteq
U_{i}$, and also have $W_{ij}\cap W_{ij'} = \emptyset$ if $m_{ij}
\neq m_{ij'}$. Therefore, if $n(u_{i+1},U_{i})\leq m_{i}/8$, then at
most $m_{i}/8$ distinct $m_{ij}$'s in $[m_{i}]$ are \emph{not} good.
Also note that the number of distinct $m_{ij}$'s \st $|W(m_{ij})| >
10|U_i|/m_i$ is less than $m_i/10$. Therefore, $|M_{i}| \geq
(\frac{3}{8} - \frac{1}{10})m_i
> m_i/4$. So if $n(u_{i+1},U_{i})\leq
m_{i}/8$, a random $m_{ij}$ in $M_i$ is good with probability at
least $1/2$, and thus $\av[J_i] \leq 2$. Also the probability that
all the Grover searches in Step \ref{step: boundary test} are
correct is at least $1- J_i\epsilon_3$.

\ref{statement: all boundary good}: We shall first prove $B(U_{i+1})
\subseteq B(U_{i}) \cup W_{i}$. In fact, any $s\in B(U_{i+1})$
satisfies that $s\in U_{i+1}$ and that $\exists t\in V-U_{i+1}$ \st
$|s-t| = 1$. Recall that $U_{i+1}\subseteq U_{i}$, so if $t\in
V-U_{i}$, then $s\in B(U_{i})$ by definition. Otherwise $t\in
U_{i}-U_{i+1}$, and thus $t\in U_{i}$ and $|t-u_{i+1}|
> m_{i+1}$ by the definition of $U_{i+1}$. Noting that
$|s-u_{i+1}| \leq m_{i+1}$ since $s\in U_{i+1}$, and that $|s-t| =
1$, we have $|s-u_{i+1}| = m_{i+1}$, which means $s\in W_{i}$. Thus
for all $s\in B(U_{i+1})$, either $s\in B(U_{i})$ or $s\in W_{i}$
holds, which implies $B(U_{i+1}) \subseteq B(U_{i})\cup W_{i}$.

Applying the result recursively, we have $B(U_{i+1}) \subseteq
B(U_{0}) \cup W_{0} \cup...\cup W_{i} = W_{0} \cup...\cup W_{i}$.
Since we have $f(u_{i+1}) \leq f(u_i) \leq ... \leq f(u_1)$ (by Step
\ref{step: point update}) and $f(u_{k+1}) \leq f(W_{k})$ (for $k =
0, ..., i$) by the assumption that all $W_k$'s are good, we know
that $f(u_{i+1}) \leq f(W_{0} \cup...\cup W_{i})$, which implies
$f(u_{i+1}) \leq f(B(U_{i+1}))$.

For the other goal $u_{i+1} \notin B(U_{i+1})$, it is sufficient to
prove $u_{i+1} \notin B(U_{i})$ and $u_{i+1} \notin W_{i}$. The
latter is easy to see by the definition of $W_i$. For the former, we
can actually prove $u_{k+1} \notin B(U_{k})$ for all $k = 0,...,i$
by induction on $k$. The base case of $k=0$ is trivial because
$B(U_0) = \emptyset$. Now suppose $u_{k} \notin B(U_{k-1})$. There
are two cases of $u_{k+1}$ by Step \ref{step: point update}. If
$f(u_k) \leq f(v_k)$, then $u_{k+1} = u_k \notin B(U_{k-1})$ by
induction. Again by the definition of $W_{k-1}$ we know that $u_{k}
\notin W_{k-1}$ and thus $u_{k+1} = u_k \notin B(U_k)$. The other
case is $f(u_k) > f(v_k)$, then $u_{k+1} = v_k$, and therefore
$f(u_{k+1}) = f(v_k) < f(u_k) \leq f(B(U_k))$ (by the first part in
\ref{statement: all boundary good}), which implies that $u_{k+1}
\notin B(U_k)$.
\end{proof}

\noindent (Continue the proof of Theorem \ref{thm: upper bound}) Now
by the claim, we know that with probability at least
$1-I(\epsilon_1+\epsilon_2) - \sum_{i=0}^{I-1}J_i\epsilon_3$, we
will have that
\begin{equation}n(u_I, U_I)\leq m_I,\qquad f(u_I)\leq
f(B(U_I)), \qquad u_I\notin B(U_I).
\end{equation}
Note that the correctness of the algorithms follows from these three
items. Actually, by the last two items, we know that any decreasing
path from $u_I$ is contained in $U_I$. Otherwise suppose $(u_I^0,
u_I^1, ..., u_I^T)$ is a decreasing path from $u_I$ (so $u_I^0 =
u_I$), and the first vertex out of $U_I$ is $u_I^t$, then $u_I^{t-1}
\in B(U_I)$. Since $u_I^0 \notin B(U_I)$, we have $t-1>0$ and thus
$f(u_I^{t-1}) < f(u_I)$, contradicting to $f(u_I)\leq f(B(U_I))$.
Now together with the first item, we know that any decreasing path
from $u_I$ is no more than $m_I$ long. Thus Step \ref{step: final
search} will find a local minimum by following a decreasing path.

The error probability of the algorithm is $I(\epsilon_1+\epsilon_2)
+ J \epsilon_3 + 10/100$, where $J = \sum_{i=0}^{I-1}J_i$. Since
$\av[J] = 2I$, we know by Markov inequality that with $J < 20I$ with
probability at least 9/10. Since $\epsilon_1 = \epsilon_2 =
1/(10\log_2 d)$ and $\epsilon_3 = 1/(200\log_2 d)$, and note that $I
\leq \log_2 d$ because $m_{0} = d$ and $m_{i+1} \leq \lceil
m_{i}/2\rceil$. So the total error probability is less than 1/2.

We now consider the number of queries used in the $i$-th iteration.
Note from Step \ref{step: initialize} and Step \ref{step: length and
range update} that $|U_{i}| \leq c(m_{i})$ for $i = 0,1,...,I-1$. So
Step \ref{step: min search} uses
\begin{equation}
O\left(\sqrt{\frac{8|U_{i}|}{m_{i}} \log \log d}\log \log d\right) =
O\left(\sqrt{\frac{c(m_{i})}{m_{i}}} (\log \log d)^{1.5}\right)
\end{equation}
queries. Also note from Step \ref{step: boundary sample} that
$|W_{ij}| \leq 10|U_i|/m_i$, so Step \ref{step: boundary search}
uses $O(\sum_{j=1}^{J_i}\sqrt{c(m_i)/m_i} \log\log d )$ queries,
which has the expectation of $O(\sqrt{c(m_i)/m_i} \log\log d )$.
Finally, Step \ref{step: final search} uses $O(\sqrt{\delta})$
queries. Note that $\delta = c(1) = O(c(m_I)/m_I)$ where $m_I$ is a
constant integer in the range $[6,10]$. Altogether, the total
expected number of queries used is
\begin{equation}
O\left(\left(\sum_{i=0}^{\log_2 d-1} \sqrt{c(m_{i})/m_{i}}\right)
(\log \log d)^{1.5} \right).
\end{equation}

If $c(k) = O(k^\alpha)$ for some $\alpha \geq 1$ and $k = 1, ...,
d$, then
\begin{eqnarray}
\sum_{i=0}^{\log_2 d-1} \sqrt{\frac{c(m_i)}{m_i}} =
\sum_{i=0}^{\log_2 d-1} m_i^{(\alpha-1)/2} = \sum_{i=0}^{\log_2 d-1}
(d/2^i)^{(\alpha-1)/2} = \frac{d^\beta - 1}{1-2^{-\beta}}
\end{eqnarray}
where $\beta = (\alpha-1)/2$. This completes the proof for the
quantum algorithm, except that in the case of $\alpha = 1$ we only
have a quantum upper bound of $O(\log d (\log \log d)^{1.5})$. But
note that the randomized algorithm uses $O(\log d \log \log d)$
queries (because of the saving at error probability controls). So
when $\alpha = 1$, the quantum algorithm just uses the randomized
one.
\end{proof}

%%%%%%%%%%%%%%%%%%%%%%%%%%%%%%%%%%%%%%%%%%%%%%%%%%%%%%%%%%%%%%%%%%%
%%%%%%%%%%%%%%%%%%%%%%%%%%%%%%%%%%%%%%%%%%%%%%%%%%%%%%%%%%%%%%%%%%%
\section{Open problems: remaining gaps}\label{sec: conclusion}
We list those grids on which the query complexities of Local Search
still have gaps.
%\begin{table}[h]
\begin{center}
\begin{tabular}{|c|c|c|}
\hline
 $d$ = & 2 & 3 \\
\hline
 old RLS & $[\Omega(1), O(n)]$ & $[\tilde{\Omega}(\sqrt{n}), O(n^{\frac{3}{2}})]$  \\
\hline
 new RLS & $[\Omega(n^{\frac{2}{3}}), O(n)]$ &  $[\Omega(n^{\frac{2}{3}}/(\log n)^{\frac{1}{2}}), O(n^{\frac{3}{2}})]$ \\
\hline
 remaining gap & $n^{\frac{1}{3}} \ (= N^{\frac{1}{6}})$ & $(\log n)^{\frac{1}{2}} \ (= (\log N)^{\frac{1}{2}})$\\
\hline
\end{tabular}
\end{center}
%\end{table}
\begin{center}
\begin{tabular}{|c|c|c|c|c|}
\hline
 $d$ = & 2 & 3 & 4 & 5\\
\hline
 old QLS & $[\Omega(n^{\frac{1}{4}}), O(n^{\frac{2}{3}})]$ & $[\tilde{\Omega}(n^{\frac{1}{4}}), O(n)]$ & $[\tilde{\Omega}(n^{\frac{1}{2}}), O(n^{\frac{4}{3}})]$ & $[\tilde{\Omega}(n^{\frac{3}{4}}), O(n^{\frac{5}{3}})]$ \\
\hline
 new QLS & $[\Omega(n^{\frac{2}{5}}), O(n^{\frac{1}{2}})]$ &  $[\Omega(n^{\frac{3}{4}}), O(n)]$ & $[\Omega(n^{\frac{6}{5}}), O(n^{\frac{4}{3}})]$ &  $[\Omega(n^{\frac{5}{3}}/(\log n)^{\frac{1}{3}}), O(n^{\frac{5}{3}})]$\\
\hline
 remaining gap & $n^{\frac{1}{10}} (= N^{\frac{1}{20}})$ & $n^{\frac{1}{4}} (= N^{\frac{1}{12}})$ & $n^{\frac{2}{15}} (= N^{\frac{1}{30}})$ & $(\log n)^{\frac{1}{3}} \ (= (\log N)^{\frac{1}{3}})$ \\
\hline
\end{tabular}
\end{center}
where $N = n^d$ is the number of the vertices in the grid.

%%%%%%%%%%%%%%%%%%%%%%%%%%%%%%%%%%%%%%%%%%%%%%%%%%%%%%%%%%%%%%%%%%%
%%%%%%%%%%%%%%%%%%%%%%%%%%%%%%%%%%%%%%%%%%%%%%%%%%%%%%%%%%%%%%%%%%%
%\section{Concluding Remarks: further improvements}\label{sec: conclusion}
%The paper gives new lower and upper bounds for Local Search
%problems.

%Random walk has been widely studied as a sampling method for
%algorithm designing, where the key parameter is the mixing time. It
%is interesting that \cite{Aa04, SS04a} and this paper use random
%walk to give lower bounds. And we can see from Theorem \ref{thm:
%general lb} that for lower bounds, we care not only about the mixing
%time of the random walk, but also about its behavior before mixing.

%%%%%%%%%%%%%%%%%%%%%%%%%%%%%%%%%%%%%%%%%%%%%%%%%%%%%%%%%%%%%%%%%%%
%%%%%%%%%%%%%%%%%%%%%%%%%%%%%%%%%%%%%%%%%%%%%%%%%%%%%%%%%%%%%%%%%%%
\vspace{3em}\noindent\textbf{Acknowledgement}

The author thanks Scott Aaronson, Xiaoming Sun and Andy Yao very
much for many valuable discussions. Thanks also to Yves Verhoeven
and Dirk Winkler for each pointing out an error in a preliminary
version of the paper.

%%%%%%%%%%%%%%%%%%%%%%%%%%%%%%%%%%%%%%%%%%%%%%%%%%%%%%%%%%%%%%%%%%%
%%%%%%%%%%%%%%%%%%%%%%%%%%%%%%%%%%%%%%%%%%%%%%%%%%%%%%%%%%%%%%%%%%%

\end{document}